\documentclass[twocolumn,sort&compress,square,numbers]{revtex4}%%%%where rsos is the template name

\usepackage{amsmath,amsfonts,amssymb}
\usepackage[english]{babel}
\usepackage[latin1]{inputenc}
\usepackage[T1]{fontenc}
\usepackage{color}
\usepackage{float}
\usepackage{verbatim}
\usepackage{graphicx}
\usepackage{bm}
\usepackage{mathtools}
\usepackage{stmaryrd}
\usepackage{anyfontsize}
\usepackage{color}
\usepackage[sectionbib]{bibunits}
\defaultbibliographystyle{amnatnat}

\defaultbibliography{aa_starving3}

\usepackage{caption}
\begin{document}

%%%% Article title to be placed here
\title{On the dynamics of mortality and the ephemeral nature of mammalian megafauna}

\author{%%%% Author details
Taran Rallings$^{1}$, Christopher P Kempes$^{2}$, Justin D. Yeakel$^{1,*}$} 
% and X. Third author$^{3}$}%

%%%%%%%%% Insert author address here
\address{
$^{1}$School of Natural Sciences, University of California Merced\\
$^{2}$Santa Fe Institute\\
$^{*}$Corresponding author: jyeakel@ucmerced.edu}
% $^{2}$Second author address\\
% $^{3}$Third author address}

%%%% Subject entries to be placed here %%%%
% \subject{foodwebs, paleontology, ecology}
%%%% Keyword entries to be placed here %%%%
% \keywords{foodwebs, mammals, foraging}

%%%% Insert corresponding author and its email address}
% \corres{T Rallings, J D Yeakel\\
% \email{trallings@ucmerced.edu}\\
% \email{jyeakel@ucmerced.edu}}

\begin{abstract}
Energy flow through consumer-resource interactions is largely determined by body size. Allometric relationships govern the dynamics of populations by impacting rates of reproduction, as well as alternative sources of mortality, which have differential impacts on smaller to larger organisms. Here we derive and investigate the timescales associated with four alternative sources of mortality for terrestrial mammals: mortality from starvation, mortality associated with aging, mortality from consumption by predators, and mortality introduced by anthropogenic subsidized harvest. The incorporation of these allometric relationships into a minimal consumer-resource model illuminates central constraints that may contribute to the structure of mammalian communities. Our framework reveals that while starvation largely impacts smaller-bodied species, the allometry of senescence is expected to be more difficult to observe. In contrast, external predation and subsidized harvest have greater impacts on the populations of larger-bodied species. Moreover, the inclusion of predation mortality reveals mass thresholds for mammalian herbivores, where dynamic instabilities may limit the feasibility of megafaunal populations. We show how these thresholds vary with alternative predator-prey mass relationships, which are not well understood within terrestrial systems. Finally, we use our framework to predict the harvest pressure required to induce mass-specific extinctions, which closely align with previous estimates of anthropogenic megafaunal exploitation in both paleontological and historical contexts.
Together our results underscore the tenuous nature of megafaunal populations, and how different sources of mortality may contribute to their ephemeral nature over evolutionary time.
\end{abstract}

\maketitle

\begin{bibunit}

\section*{Introduction}

% [P1: consumer-resource systems and allometry... energy compartments...paleo]
Consumer-resource interactions are the fundamental unit from which complex food webs arise \cite{deangelis1980energy}. In such dynamics, the rates governing transitions of biomass and energy from one species to another are largely determined by body size \cite{Yodzis:1992hg}. 
Specifically, the allometric relationships between consumer body mass and metabolic rate constrain energetic assimilation \cite{hou2008}, storage \cite{Lindstedt:2002td}, and growth \cite{West:2001bv}, all of which  govern the dynamics of populations \cite{hennemann1983relationship,West:2002it,Kempes:2012hy,yeakel2018dynamics}. 
% The scaling of the ratio of fat to non-fat mass in mammals, for example, is central to understanding how these physiological constraints influence population level dynamics. 
% This ratio of fat to non-fat mass is the relative weight of energy storage to metabolism in an individual. 
% The ratio scales predictably and positively with body-mass which creates size-structured patterns like starvation tolerance. 
Because allometrically-constrained models of population dynamics apply generally across large taxonomic clades, they are useful for examining dynamic constraints that may contribute to community structure across macroevolutionary timescales \cite{DeLong:2012fjb,DeLong2012carnivores,Pawar2012,yeakel2018dynamics,bhat2020scaling}.
Furthermore, examination of community dynamics at these scales enables the investigation of extinct communities where body size distributions were different than those in contemporary ecosystems \cite{alroy2001multispecies,brook2008synergies,bradshaw2021relative}.
% The constraints imposed by these physiological allometries on population dynamics allow us to make predictions about populations that we cannot directly observe. 
% This is notably important for understanding paleo-ecosystems as our knowledge is often limited to body-mass and taxa. 
% By combining energetic models with allometric parameters we can approximate the energy needs and use of extinct consumers and interrogate sized-structured ecological patterns such as the megafaunal extinction at the terminal Pleistocene. 

% [P2: different sources of mortality]
The dynamics of populations represent an energetic balance between reproduction and mortality \cite{murdoch:2003}.
Across Mammalia, the average rate of reproduction can be predicted from allometric scaling relationships \cite{CalderIII:1983jd}, though individual clades demonstrate a large variety of reproductive strategies -- from changing reproductive cycles, litter sizes, and dynamic responses to changing resource conditions to name a few \cite{roff1993evolution}.
As these strategies are typically evolved responses to particular conditions and clade-specific, they are not universally experienced.
On the other hand, mortality has a variety of forms that nearly all species must deal with to a greater or lesser extent, and do not all scale similarly with body size \cite{weitz2006size}.
Mortality originates from both internal and external drivers, where the former depends on an organism's internal state to initiate death.
For example, senescence and starvation involve physiological states that change with respect to clock time, metabolic rate, and resource depletion \cite{yeakel2018dynamics,robert2015actuarial}.
In contrast, external drivers of mortality consist of an outside force that induces death more independently of an organism's internal state, such as mortality due to natural predation or subsidized anthropogenic harvest.
Often, mortality occurs through correlations between internal and external drivers, where for example, the starvation state of prey may alter the success rates of predators \cite{Alonzo:2002ub}.
% And while we consider these internal and external drivers separately here, they represent a continuum where mortality occurs due to 
While virtually all primary consumer populations must deal with the effects of resource limitation, aging, and predation, the effects of anthropogenic harvesting (the subsidized extraction of prey) are uniquely limited to those species serving as resources for human populations \cite{Dunne:2016jp}.
% on terrestrial mammals during the end-Pleistocene and throughout the Holocene are thought to have resulted in considerable community restructuring \cite{Koch:2006vt,Estes:2011eo,Yeakel2014}.

% [P3: constraints of size]
How do different sources of mortality impact the dynamics of mammalian populations?
Here we construct a general consumer-resource framework to examine mammalian herbivore populations as a function of consumer body size $M_C$, as well as size-dependent vulnerability to different internal and external pressures.
Our approach integrates relationships governing specific timescales of physiology and assimilation from a process-based energetic perspective \cite{West:2001bv}. 
Our model is low-dimensional and compact \cite[cf.][]{yeakel2018dynamics}, but due to its close connection to fundamental energetic mechanism, it is also capable of reproducing observed large-scale empirical patterns of mammalian communities.
We begin by describing our approach, reproducing key macroecological relationships such as Damuth's law \cite{Damuth1987}, and examine how changes to energetic parameters impact these predictions.
We then derive timescales associated with four sources of mortality experienced by mammalian consumers: \emph{i}) natural mortality, \emph{ii}) starvation mortality, \emph{iii}) natural predation, and \emph{iv}) subsidized anthropogenic harvest.
By examining each source of mortality in turn, our framework illuminates central constraints governing mass-specific behaviors, strategies, and risks experienced by mammalian consumers.

Our results reveal four key insights into the constraints structuring mammalian communities.
First, our allometric consumer-resource system accurately captures both the central tendency and variability of Damuth's law, suggesting that the included vital rates capture mass-specific dynamics.
Second, our results demonstrate that natural and starvation mortality differentially impact small mammals, confirming expectations, and point to why the allometric effects of senescence are difficult to observe in nature.
Third, we detail the differences in how mortality under different levels of predation intensity induce dynamic instabilities for large-bodied megaherbivores. 
We also show that the body size at which these instabilities occur is dependent on the prevailing predator-prey mass relationship (PPMR).
Finally, we evaluate the harvest pressure required to induce mass-specific extinction, and show that our predictions are comparable to estimates of both paleontological and historical exploitation of mammalian megafauna.

\section*{Methods}

We model a consumer-resource interaction, where the resource $R$ (g/m${}^2$) grows logistically with intrinsic growth rate $\alpha$ to a carrying capacity $k$, and declines due to consumption by an herbivore consumer population $C$ (g/m${}^2$) (Eq. \ref{eq:2d}).
Consumed resources govern both consumer somatic maintenance and reproduction. 
The rate of consumption to fuel somatic maintenance is given by $\rho$, and is independent of resource density, as these are invariant requirements of the consumer population \cite{yeakel2018dynamics}.
In contrast, the rate of consumption to fuel reproduction is proportional to resource density and is given by $\lambda_C(R)/Y_C$, where $\lambda_{\rm C}(R)$ is the consumer growth rate and $Y_{\rm C}$ is the consumer yield coefficient, or the grams of consumer produced per gram of resource consumed.
% We assume the consumer growth rate
% , and the consumer yield coefficient describes the grams of consumer produced per gram of resource consumed (see Methods).
As in DeLong \& Vasseur \cite{DeLong:2012fjb}, the consumer's growth rate $\lambda_C(R)$ follows a Type II (saturating) functional response given the resource density $R$, where the maximum growth is $\lambda^{\rm max}_{\rm C}$ and the resource half-saturation density is $\hat{k} = k/2$, such that 
\begin{equation}
    \lambda_{\rm C}(R) = \lambda^{\rm max}_{\rm C}\left(\frac{R}{\hat{k}+R}\right).
    \label{eq:growth}
\end{equation}
% As the resource diminishes, consumer reproduction is slowed.

While the consumer population density grows at rate $\lambda_{\rm C}(R)$, we assume for now that consumer mortality is a function of both natural mortality $\mu$ and starvation $\sigma(R)$, where the rate of starvation,
\begin{equation}
\sigma(R) = \sigma^{\rm max}\left(1-\frac{R}{k}\right),
\label{eq:starve}
\end{equation}
increases as resources become scarce.
In this context, $\sigma^{\rm max}$ is the maximal rate of starvation that occurs when the environment is devoid of resources.
The full system describing resource and consumer dynamics is given by
\begin{align}
    \frac{{\rm d}}{{\rm dt}}C &= \lambda_{\rm C}(R)C - \left(\mu + \sigma(R) +...\right)C, \nonumber \\ 
    \frac{{\rm d}}{{\rm dt}}R &= \alpha R \left(1 - \frac{R}{k} \right) - \left(\frac{\lambda_{\rm C}(R)}{Y_{\rm C}} + \rho \right)C,
    \label{eq:2d}
\end{align}
where the `$...$' denotes where additional mortality terms, described later, will be included.
The dynamic outcomes of this system of equations include two trivial steady states at $(R^*=0, C^*=0)$ and $(R^*=k,C^*=0)$, and one internal steady state where both the consumer and resource population coexist.
% Because the internal steady state cannot be concisely written, we do not report it here.
See tab. \ref{tab:param} for a description of parameters.
% \begin{align}
%     C^* &= \alpha k\frac{2\lambda^{\rm max}_{\rm C}\sigma Y_{\rm C}}{(2\lambda^{\rm max}_{\rm C}+\sigma)(2\lambda^{\rm max}_{\rm C} \sigma + (2\lambda^{\rm max}_{\rm C}+\sigma)Y_{\rm C}\rho)} \nonumber \\ 
%     R^* &= k \frac{\sigma}{2\lambda^{\rm max}_{\rm C} + \sigma},
%     \label{eq:2dss}
% \end{align}
% where both the consumer and resource population coexist.

The rate laws describing resource consumption as well as consumer growth and mortality all vary as a function of consumer body mass $M_C$, where the consumer is assumed to be a mammalian herbivore, and the resource is an unspecified primary producer with characteristic growth rate, carrying capacity, and energy density $E_d$. %(see methods for derivations of allometric rate laws).
% While we describe the detailed derivations of allometric rate laws in methods (see methods), it is useful to briefly summarize how consumer rates scale with mass.
We approach the derivation of vital rates with respect to consumer mass by solving for multiple timescales associated with ontogenetic growth, maintenance, and expenditure.
The growth of an individual consumer from birth mass $m=m_0$ to its reproductive size $m=0.95M_C$ is given by the solution to the general balance condition $B_0 m^\eta = E_m \frac{\rm d}{\rm dt}m + B_m m$, where $E_m$ is the energy needed to synthesize a unit of biomass, $B_m$ is the metabolic rate to support an existing unit of biomass (tab. \ref{tab:param}), and the metabolic exponent $\eta=3/4$ \cite{Pirt1965,West:2001bv,hou2008}.
% IF WE HAVE TO CUT THE REST TO AN APPENDIX, WE COULD...
From this balance condition, the time required for an organism starting from mass $m_1$ to reach mass $m_2$ follows
\begin{equation}
    \tau(m_1,m_2) = \ln\left(\frac{1 - (m_1/M_C)^{1-\eta}}{1-(m_2/M_C)^{1-\eta}}\right)\frac{M_C^{1-\eta}}{a(1-\eta)}
    \label{eq:timescale}
\end{equation}
where $a = B_0/E_m$ \cite{West:2001bv}.
We use this general timescale equation to calculate maximal rates of growth and starvation as a function of organismal body size $M_C$, which are then modified by resource density $R$ to provide realized timescales (eqs. \ref{eq:growth},\ref{eq:starve}).
We note that a more complex framework could include the effects of changing resource densities on timescales directly, where individual growth is itself variable, effectively introducing dynamic population structure \cite{deRoos2020}. 
From this general equation, we calculate the timescale of reproduction for an herbivore consumer of mass $M_C$ as $t_{\lambda_C} = \tau(m_0,0.95M_C)$, such that the maximal reproductive rate is $\lambda^{\rm max}_C = \ln(\nu)/t_{\lambda_C}$, where $\nu=2$ is the set number of offspring per reproductive cycle \cite{Savage2004,yeakel2018dynamics}.
The consumer yield coefficient is given by $Y_C = M_C E_d/B_{\lambda_C}$ (g consumer per g resource), where $B_{\lambda_C}$ is the lifetime energy use required by the herbivore to reach maturity $B_{\lambda_C} = \int_0^{t_{\lambda_C}} B_0 m(t)^\eta {\rm d}t$,
and the maintenance rate is given by $\rho = B_0 M_C^\eta/M_C E_d$ \cite{yeakel2018dynamics}.

To determine the rate of mortality from starvation, we calculate the time required for an organism to metabolize its endogenous energetic stores, estimated from its cumulative fat and muscle mass, where the remaining mass is given by $M_C^{\rm starve} = M_C - (M_C^{\rm fat} + M_C^{\rm musc})$ (see app. C).
During starvation, we assume that an organism burns its existing endogenous stores as its sole energy source, such that the balance condition is altered to $\frac{\rm d}{\rm dt}m E^\prime_m = -B_m m$, where $E_m^\prime$ is the amount of energy stored in a unit of biomass (differing from the amount of energy used to synthesize a unit of biomass $E_m$; tab. \ref{tab:param}) \cite{moses2008rmo,hou2008}.
The starvation timescale is then given by
\begin{equation}
    t_{\sigma} = -\frac{M_C^{1-\eta}}{a^\prime}\ln(M_C^{\rm starve}/M_C), %^{\rm max}
\end{equation}
where $a^\prime = B_0/E^\prime_m$, such that the starvation rate is the $\sigma^{\rm max} = 1/t_{\sigma}$. %^{\rm max}
Importantly, the starvation mortality expressed here is specifically that experienced by adult organisms (as the timescale of metabolizing fat stores is conditioned on adult mass), and does not capture potential starvation mortality of juveniles.
% Two additional allometric constants are required to populate Eq. \ref{eq:2d}, including the maintenance rate $\rho$ and the yield coefficient $Y_C$.

To determine the rate of mortality from aging, we note that population cohorts experience two primary sources of natural mortality: the initial cohort mortality rate $q_0$ and the annual rate of increase in mortality as the cohort ages, or the actuarial aging rate, $q_a$ over lifetime $t_\ell$.
We begin by assuming that the number of survivors over time follows a Gompertz relationship \cite{CalderIII:1983jd} from which we derive the average rate of natural mortality 
\begin{equation}
    \mu = \frac{q_0}{q_a t_\ell}\big({\rm exp}(q_a t_\ell)-1\big).
\end{equation}
% and where $\mu$ influences consumer population densities as
% \begin{equation}
%     \frac{{\rm d}}{{\rm d}t}C = \lambda^{\rm max}_{\rm C}RC- \left(\sigma \left(1 - \frac{R}{k}\right) + \mu \right)C.
% \end{equation}
The three parameters $(q_0,q_a,t_\ell)$ each have well-documented allometric relationships for terrestrial mammals, such that natural mortality can be written as a function of consumer mass $\mu(M_C)$ (see app. A).
Because both cohort and actuarial mortality are not subdivided into specific categories, we may assume that this rate is capturing the combined effects of all sources of mortality, particularly during early development when starvation and predation risks are highest.

% Finally, the maintenance rate is given by $\rho = B_0 M_C^\eta/M_C E_d$.
% We emphasize that while 
As the sizes of physiological biomass compartments are obtained from empirical observations, the rates determining biomass flux are derived from process-based energetic relationships (eq. \ref{eq:timescale}).
Together, the allometric rate laws and the dynamic system presented in Eq. \ref{eq:2d} allow us to assess the dynamics of consumer-resource systems for mammalian herbivores spanning the observed range of terrestrial body sizes, from the smallest (the Etruscan shrew at roughly $1$ g) to the largest (the Oligocene paraceratheres and Miocene deinotheres at ca. $1.5-1.74\times10^7$ g) \cite{Smith:2010p3442}.
We next examine how this minimal framework is well-suited to provide general insight into several key allometric constraints that contribute to the functioning and limitations of terrestrial mammalian communities.

\section*{Results \& Discussion}
%Comparison to NSM
\subsection*{Recovering Damuth's mass-density relationship}
% \noindent {\bf Predicting Damuth's Law} 
Our consumer-resource system is related to the nutritional state model (NSM) proposed in \citet{yeakel2018dynamics}, where an explicit starvation dynamic was incorporated by separating the consumer population density into `full' and `hungry' states. 
Here we eliminate the transition between these states, and because the timescales of transitioning between full and hungry states are short relative to those of reproduction, have sacrificed only a modest degree of physiological realism to enable analytical expression of steady states with additional sources of mortality.
If we ignore the negligible effects of $\rho$ (see app. B), analytical expression of the consumer steady state as a function of mass -- or the mass-density relationship -- follows
\begin{equation}
C^*(M_C) \approx \alpha k Y_C\frac{\sigma^{\rm max} - 2 \lambda_C^{\rm max} + A}{4 {\sigma^{\rm max}}^2},
\label{eq:steadystate}
\end{equation}
where $A = \sqrt{8{\sigma^{\rm max}}^2+(\sigma^{\rm max}-2\lambda_C^{\rm max})}$, where $\lambda_C^{\rm max}$, $\sigma^{\rm max}$ and $Y_C$ are functions of mass $M_C$.

\begin{figure}
    \centering
    \includegraphics[width=1\linewidth]{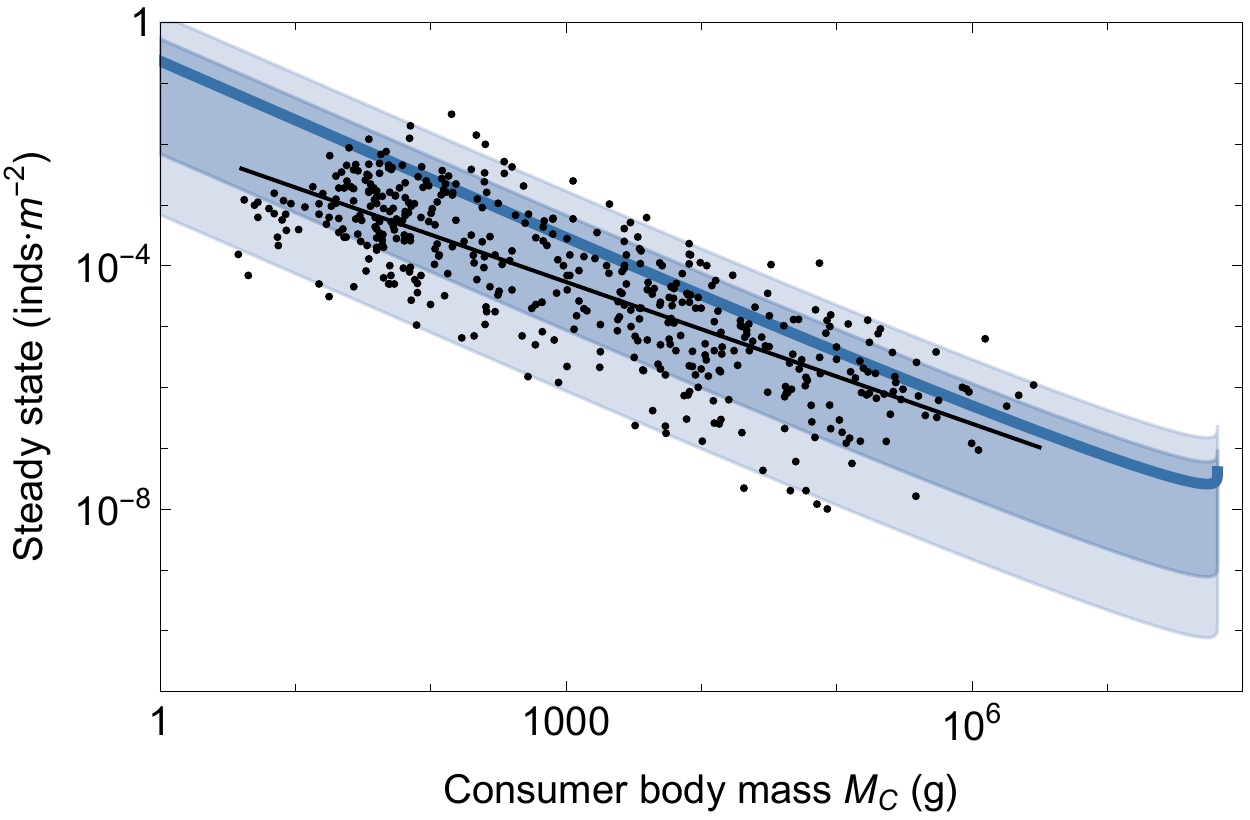}
    \caption{Model predictions of mammalian steady states ($\rm inds\cdot m^{-2}$) as a function of herbivore consumer body mass $M_C$ (thick blue line) compared to observational data from Damuth \cite{Damuth1987} (black points).
    The black line denotes the best-fit linear regression on observed densities based on ordinary least squares.
    Variation in steady state densities is captured by allowing the plant resource growth rate to vary as $\alpha = 2.81\times 10^{-10}: 2.19\times 10^{-8}~{\rm s}^{-1}$ (dark blue shaded region), and both $\alpha$ and the plant resource carrying capacity to vary as $k = 2.3:34~{\rm kg/m}{}^2$ (light blue shaded region).
    % See appendix B for details.
    % We assume an intrinsic growth rate roughly that of grass where $\alpha = 9.45\times 10^{-9}$ (s${}^{-1}$; REF), whereas observations among terrestrial plants reveal a range in growth rates from $2.81\times 10^{-10}$ to $2.19\times 10^{-8}$ \cite{michaletz2014convergence}, according with a change in $\alpha$ of roughly 97\% lower and 130\% higher than the set value. 
    % By incorporating this range into our the estimated resource growth rate, we observe that we can account for a large portion of consumer steady state densities around the mean density given by Eq. \ref{eq:2d} (inner shaded region, fig. \ref{fig:2ddensities}).
    % If we additionally adjust the carrying capacity $k$ of the resource to 90\% less-than and 150\% more-than the assumed value of $23\times 10^3$ g/m${}^2$, our framework accounts for nearly the full range of mammalian steady state densities (outer shaded region, fig. \ref{fig:2ddensities}).
    % In this context, the upper-boundary of $k$ observed to capture most higher herbivore densities is ca. 34 kg/m${}^2$, which is on the higher end of estimated live above-ground biomass densities in terrestrial forests such as in Isle Royal and the Allegheny National Forest \cite{de2017simulating}
    }
    \label{fig:2ddensities}
    \end{figure}

% , which has the effect of steepening the slope to -0.88 compared to the empirical estimate of ca. -0.77.
% We suggest that the advantages of avoiding  
% As observed in Yeakel et al. \cite{yeakel2018dynamics}, a large body mass asymptote appears at $M_C = XX$, which corresponds to the unrealistic size at which 
% The allometric consumer-resource model provides an approximation to the mass-density relationship observed among terrestrial herbivorous mammals \cite{Damuth1987}.
% The magnitude and slope of this approximation are thus products of the vital rates in Eq. \ref{eq:2d} that are driven by both physiology and the environment.
% , and it is useful to understand which components drive different aspects of the relationship.
% To disentangle the influence of the various parameters in the consumer resource model, we evaluate the influence of a percent-change to each.
% The parameters influencing just the mass-density intercept are those that determine resource availability, including the resource's intrinsic growth rate $\alpha$ and carrying capacity $k$.
The scaling of mammalian population densities was originally observed by \citet{Damuth1981,Damuth1981b} as the reciprocal of energy use requirements with an exponent of ca. -3/4.
% t \cite{Damuth1981}.
Consumer-resource models parameterized using allometric relationships can effectively predict this mass-density relationship \cite{Yodzis:1992hg,DeLong:2012fjb,yeakel2018dynamics}, while the addition of predator-prey size ratios and consumer capture relationships enable similar predictions at higher trophic levels \cite{weitz2006size,DeLong2012carnivores}.
% More recently, \citet{DeLong:2012fjb} showed that minimal population models could capture this relationship
By integrating dimensional scaling into search and consumption rates, \citet{Pawar2012} captured the mass-density relationship while highlighting potential instabilities arising in 3-Dimensional (aquatic) environments.
Our approach differs from most prior efforts by deriving timescales associated with reproduction and mortality directly from the energetic trade-offs associated with somatic growth and maintenance.
After substituting allometric relationships into the rate laws in Eq. \ref{eq:2d}, we observe that the internal steady state of consumer densities in our framework is very close, though slightly elevated, to observed mammalian densities, similarly approximating Damuth's Law (blue line, fig. \ref{fig:2ddensities}).

Our predicted mass-density relationship is premised on the assumption of resource growth rates and carrying capacities characteristic of grasses (tab. \ref{tab:param}), contributing to the slightly elevated mass-density relationship compared to the observed best-fit (black line, fig. \ref{fig:2ddensities}).
As the resource growth rate and carrying capacity are in the numerator of eq. \ref{eq:steadystate}, they determine the intercept of the relationship such that lower values will more closely match observed densities.
Along these lines, incorporating observed ranges of $\alpha$ and $k$ reveal strong alignment between model predictions and the variability of empirical mammalian densities (fig. \ref{fig:2ddensities}; see app. B for details).
Compared to the NSM \cite{yeakel2018dynamics}, and similar to \citet{DeLong2012carnivores}, our prediction reveals exaggerated densities for smaller-bodied consumers, though within the observed range of variation, resulting in a predicted mass-density relationship with a steeper slope than expected.
An elevated mass-density slope is not observed when explicit starvation and recovery are included \cite{yeakel2018dynamics}, suggesting these dynamics play an important role in depressing the populations of smaller-bodied species, in particular.
% The predicted mass-density relationship for smaller-bodied organisms is steeper ($\propto M^{0.91}$).
% \citet{Pawar2012} showed that integrating dimensional scaling into search and consumption rates captures the mass-density scaling for 2-D systems.

While eq. \ref{eq:steadystate} cannot be readily expressed when allometric relationships are included, for larger body sizes the maximal starvation rate $\sigma^{\rm max}\propto M_C^{-0.3}$, the yield coefficient $Y_C \propto M_C^{-1/4}$, the maximal consumer growth rate $\lambda_C^{\rm max} \propto M_C^{-1/4}$, and the quantity $A\propto M_C^{-0.37}$.
% (where the symbol $\appropto$ is read `approximately proportional to')
For larger body masses, this results in a predicted mass-density relationship 
% $\appropto M^{-0.76}$ (inds/m${}^2$), with a final measured value 
$\propto M_C^{-0.82}$ inds/m${}^2$, only slightly steeper than Damuth's mass-density relationship $\propto M_C^{-0.77}$ inds/m${}^2$.
At unrealistically large body sizes, the consumer steady state encounters a vertical asymptote \cite[also noted in][]{yeakel2018dynamics}.
In this region, the superlinear body fat allometry (tab. \ref{tab:param}) predicts the organism to be 100\% fat, such that the starvation timescale is infinite.
While this is mathematically entertaining, we restrict our interpretations to realistic body size ranges, thereby avoiding this particular physiological singularity.
% the difference  due to explicit starvation-recovery dynamics, which we do not include.
% which is slightly steeper than Damuth's mass-density relationship.
% We thus observe that the approximate starvation dynamics steepen the mass-density relationship slightly from that observed in the NSM \cite{yeakel2018dynamics}.
We examine additional effects of altered vital rates on the slope and intercept of the mass-density relationship in app. B. 

\subsection*{Senescence and starvation have a larger impact on smaller consumers}

% \subsection*{Initial cohort and actuarial mortality disproportionately impact smaller consumers}
% \subsection*{The Many Faces of Death} %This might be just for us ;)
% \subsubsection{Survivorship mortality (intrinsic, mass-specific, small-size punishing))}

% We consider three additional sources of mortality that are unrelated to starvation due to resource limitation: \emph{i}) initial cohort and actuarial mortality, \emph{ii}) mortality due to natural predation, and \emph{iii}) mortality due to external harvesting.
% Because these sources of mortality do not feed back to influence resource dynamics except through changes to consumer population densities, we leave the resource dynamic ${\rm d}R/{\rm d}t$ reported in Eq. \ref{eq:2d} unaltered and describe only alterations to the consumer equation.
%What is the rate of life history mortality?

\begin{figure}
    \centering
    \includegraphics[width=1\linewidth]{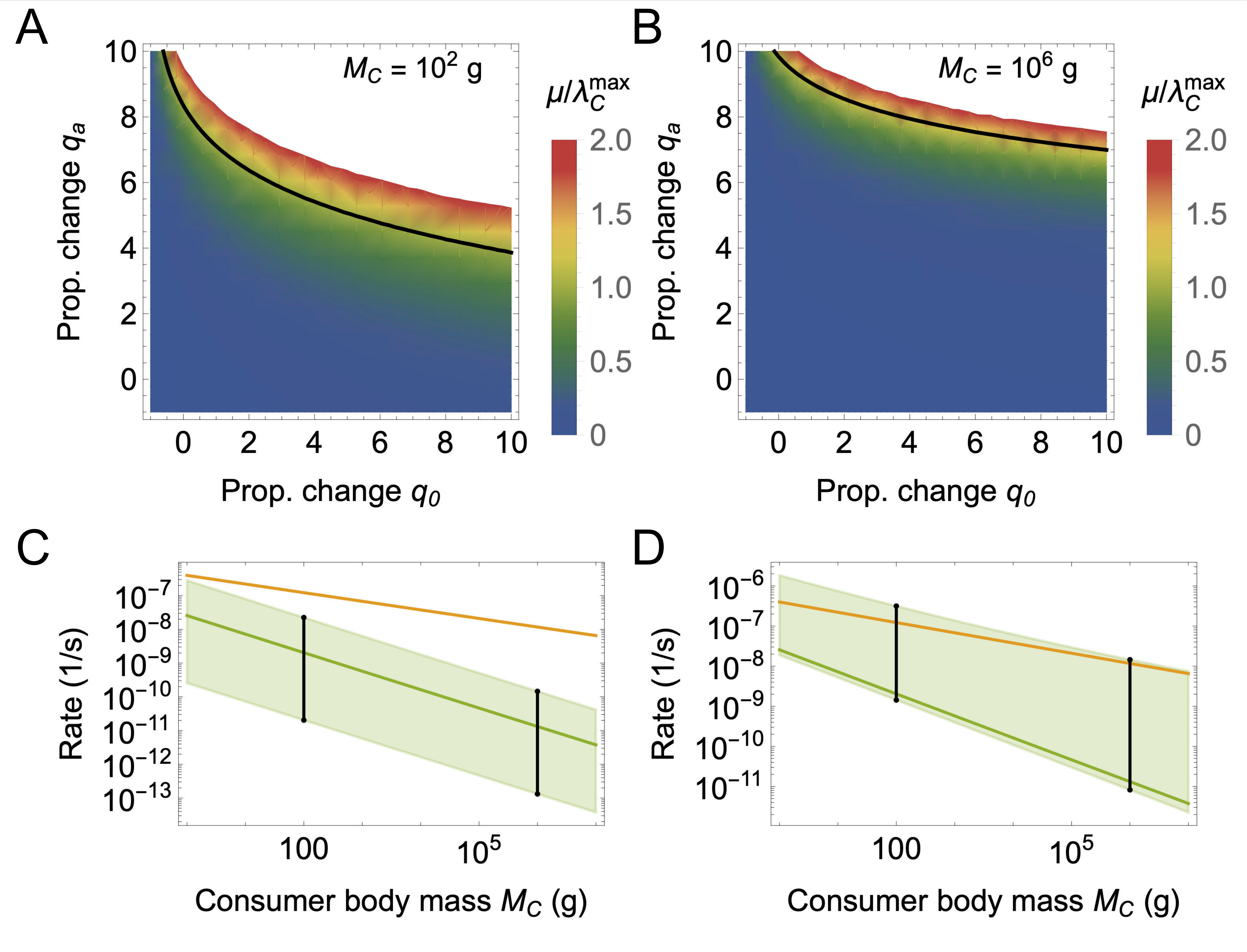}
    \caption{
        Changes in natural mortality as a function of initial cohort mortality $q_0$, and actuarial aging, $q_a$ for two different consumer body masses, $M_C$.
        A,B. The ratio reproduction $\lambda_C^{\rm max}$ to natural mortality $\mu$ for a mammalian herbivore of A. $M_C=10^2$ g and B. $M_C=10^6$ g, across proportional changes to the initial cohort mortality rate $q_0$ and the actuarial aging rate $q_a$. 
        The black contour denotes $\mu/\lambda = 1$.
        % B. As in A., but for a mammalian herbivore of $M_C=10^6$ g.
        C,D. Natural mortality $\mu$ (green) relative to reproduction $\lambda_C^{\rm max}$ (orange) as a function of consumer body mass $M_C$. The range of variation (light green shaded region) shows proportional changes to the C. cohort mortality rate $q_0$ and the D. actuarial aging rate $q_a$ from -0.99 to 10.
        % D. As in C. but with the shaded region shows proportional changes to the actuarial mortality rate $q_a$ from -0.99 to 10.
    }
    \label{fig:naturalmort}
\end{figure}

We first consider two internal sources of mortality: that due to the effects of aging, where mortality changes with an organism's temporal state, and that due to starvation, where mortality scales with an organism's energetic state.
% An immediate insight into dynamic constraints on populations can be gained by examining how the rates of consumer mortality compare to reproduction, with the expectation that stability requires mortality rates to be lower than reproductive rates (fig. \ref{fig:naturalmort}a).
% We observe that natural mortality, which accounts for both initial cohort mortality and actuarial mortality, is much lower and projects a steeper slope over $M_C$ than the maximal reproductive rate (fig. \ref{fig:naturalmort}b), with a scaling $\propto M_C^{-0.56}$ \cite[cf.][]{jones2008senescence}.
% While the calculated value of $\mu$ is nearly one order of magnitude below the rate of reproduction $\lambda_C^{\rm max}$, a steeper slope implies that increases in $\mu$ will disproportionately harm smaller organisms. 
To understand the effect of changes to $\mu(M_C)$ on consumer steady states, we examine variations in the principle components of $\mu$: initial cohort mortality $q_0$ and actuarial mortality $q_a$.
%initial cohort mortality
The initial cohort mortality represents the mortality experienced by a cohort prior to accruing effects of age. 
We observe that the mortality rate changes proportionally with $q_0$ independent of consumer mass, where the ratio $\mu/\lambda_C^{\rm max} < 1$ even with respect to large increases in $q_0$, unless $q_a$ is similarly magnified (fig. \ref{fig:naturalmort}A,B).
For survivorship mortality to approach the rate of reproduction ($\mu/\lambda_C^{\rm max}=1$), where perceptible declines in population densities result, the initial cohort mortality must increase by roughly an order of magnitude (shaded region in fig. \ref{fig:naturalmort}C).
Due to the steepness of the scaling of $\mu$ relative to $\lambda_C^{\rm max}$, this effect is felt exclusively by small-bodied organisms.

%actuarial mortality
Actuarial mortality represents the cumulative effects of aging, or senescence, across the organism's expected lifetime. 
We observe that as $q_a$ increases, the magnitude of mortality increases disproportionately (fig. \ref{fig:naturalmort}A,B), while the slope of $\mu(M_C)$ becomes more shallow (fig. \ref{fig:naturalmort}D), primarily due to the cumulative nature of senescence magnifying its effects across the longer lifetimes of larger mammals.
As such, an increase in $q_a$ overwhelms reproduction such that $\mu/\lambda_C^{\rm max} > 1$, resulting in population instability (fig. \ref{fig:naturalmort}A,B).
% destabilization of their respective populations. 
% Were an ecological context to arrive that dialed up the actuarial rate the community would be defaunated by size class from small to large.
The extinction risk imposed by senescence has been explored across mammalian taxa, and while some life-history characteristics such as the inter-birth interval appear to correlate strongly with these risks, the role of body size is notably ambiguous \cite{robert2015actuarial}.
% longer intervals between litters may promote declining population viability, while 
Though our model -- which considers averaged effects across terrestrial mammals -- predicts that the risks of increased actuarial mortality are disproportionately felt by smaller size-classes, we also show that $\mu(M_C)$ increasingly resembles $\lambda_C^{\rm max}(M_C)$ with increasing $q_a$ (the top border of the shaded region in fig. \ref{fig:naturalmort}D). % predicts that the most extreme risks imposed by increased actuarial mortality impacts smaller size-classes, 
This increased similarity implies that relatively small variations in other demographic processes or interactions may have potentially large and destabilizing effects on population size that cannot be predicted from body mass, a potential source for the noted ambiguity between size and actuarial extinction risk \cite{robert2015actuarial}.
% We suggest that such effects may be a contributing factor to the noted ambiguity in the role of body size on extinction risks due to senescence across mammalian groups \cite{robert2015actuarial}.
% Our prediction of this increased similarity between reproductive and natural mortality rates across body size may be a contributing factor for the noted ambiguity in the effects of body size on increased rates of senescence across different mammalian groups \cite{robert2015actuarial}.

\begin{figure}
    \centering
    \includegraphics[width=1\linewidth]{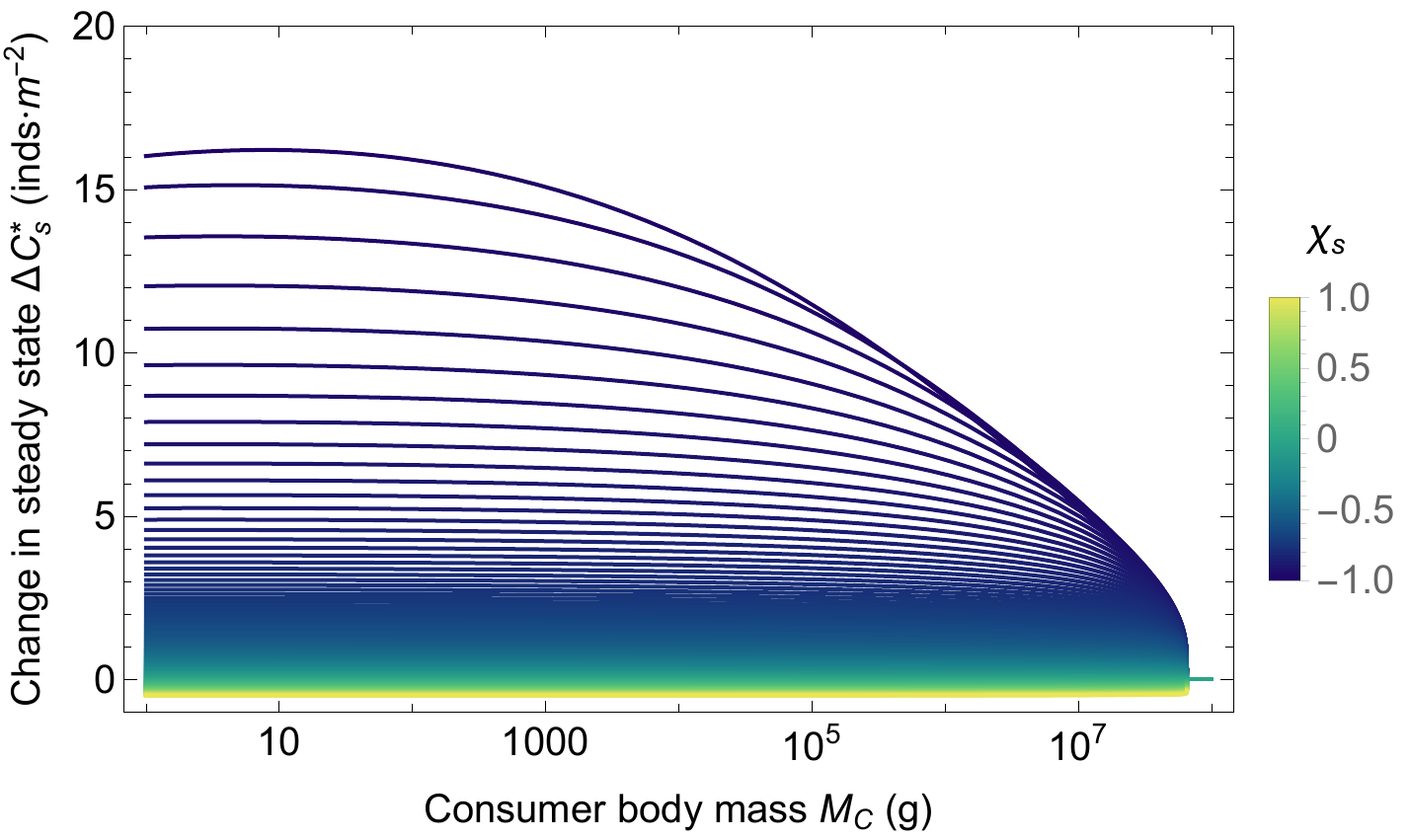}
    \caption{
    The relative change in consumer steady state $\Delta C^*_s$ as a function of consumer body mass $M_C$ given an altered rate of starvation $\sigma(R)\cdot(1 +\chi_s)$ across the proportional change $\chi_s \in (-0.99,1)$.
    }
    \label{fig:starve}
\end{figure}

% [Body mass more complicated, but here we are looking at only terrestrial organisms]
% [taken across species and not including the details of life history, while our model forecasts population crashes for small-bodies species, the shallowing of $\mu$ with increasing actuarial mortality results in a very similar $\mu$ and $\lambda_C$ across $M$]
% [This means small variations in other life history variables may have unpredictable effects on population size]
% [explains some of the ambiguity in the effects of actuarial mortality as a function of body size.]

While the temporal state of an organism is unidirectional and linear, other internal states, such as an organism's energetic state, fluctuate nonlinearly over time.
In this case, the rate of starvation is low when resources become plentiful ($R \rightarrow k$) and increases to $\sigma^{\rm max}$ as resources become scarce ($R \rightarrow 0$).
Because organisms metabolize their fat and muscle tissue during starvation, and die from starvation when these energetic stores are metabolized, the timescale of starvation varies with the amount of endogenous energetic stores an organism carries.
Larger organisms carry a larger proportion of body mass as fat \cite{Lindstedt:2002td}, such that they are more protected from the effects of short-term resource scarcity \cite{Millar:1990p923}.
% resulting in a strong allometric relationship for mortality from starvation \cite{yeakel2018dynamics}.
We observe this effect by modifying the starvation rate and examining how the steady state population size is altered.
% Variations to starvation-induced mortality have a large effect on consumer steady states, though the effect is not the same across body mass.
We introduce variation to the rate of starvation as $\sigma(R)\cdot(1+\chi_s)$, from which the altered steady state $C^*_{s}$ is calculated.
The relative change in steady states introduced by the altered starvation rate is then given by $\Delta C^*_s = (C^*_{s} - C^*)/C^*$,
% introduced by the modified rate is calculated as 
% \begin{equation}
% \Delta C^*_s = \frac{C^*(M_C|\sigma(R)\cdot(1+\chi_s)) - C^*(M_C|\sigma(R))}{C^*(M_C|\sigma(R))},
% \label{eq:starve}
% \end{equation}
where positive values indicate a relative gain in steady state densities from the proportional change $\chi_s$, and negative values indicate a relative loss (fig. \ref{fig:starve}).
We observe that, while all mammals benefit from reduced starvation rates ($\chi_s<0$), smaller-bodied mammals benefit to a much greater extent, and this effect tapers off with increasing body mass.
Because fat biomass scales super-linearly with body mass, the populations of larger consumers are more resilient to the effects of starvation, whereas those of smaller consumers are more prone.

An organism's rate of starvation emerges from two governing forces -- the amount of energy storage and the rate of its use -- and as such can be be manipulated both physiologically and behaviorally.
For instance, behaviorally supplementing endogenous fat stores with exogenous caches magnifies an individual's energetic stores \cite{Lucas1991,yeakel2020caching}, whereas physiologically-mediated responses to starvation risk such as torpor can introduce significant temporal delays to the effects of resource scarcity \cite{schubert2010daily}.
% While these mechanisms are very different, 
In both cases the time required to pass from a replenished to a starved state is effectively increased, lowering the rate of starvation.
% As shown in fig. \ref{fig:corr}E,F, 
The predicted benefits of such adaptations to mammalian steady state densities will be realized primarily by smaller mammals (fig. \ref{fig:starve}, app. B), and it is the smaller body size range where traits such as caching and torpor are most commonly observed \cite{geiser1998evolution,smith1984evolution,yeakel2020caching}. 

% Additional sources of mortality may introduce very different effects on consumer populations, and we next explore how mortality beyond starvation informs our understanding of size-dependent extinction risks.
% We next examine the effects of \emph{i}) initial cohort and actuarial mortality, \emph{ii}) mortality due to natural predation, and \emph{iii}) mortality due to external harvesting.

\subsection*{Predation mortality and the feasibility of megatrophic interactions}
% (extrinsic, mass-specific, small-size punishing)}

Predators introduce an external source of mortality on prey populations, fueling their own population growth in whole (trophic specialists) or in part (trophic generalists), by the rate at which prey are consumed.
% We next explore the effect of predator mortality on an herbivore consumer, taking into account carnivore-specific growth and energetic rates.
% Without explicitly taking into account the dynamics of a predator population, 
We account for the effects of an implicit predator density $P$ with body size $M_P$ on the herbivore consumer density $C$ with body size $M_C$, where we assume the predator population to exist at a fixed density $P \equiv P^*$.
% , where we will for now assume that the predator population exists at some steady state $P^*$.
% Using a form similar to consumer-resource consumption in Eq. \ref{eq:2d}, 
The mortality rate of the herbivore consumer from an external predator is given by 
\begin{equation}
\beta(C,P)=w\frac{\lambda_P(C) P}{C Y_P}, 
\end{equation}
where $\lambda_P(C)$ is the growth rate of the predator and $Y_P$ is the predator yield coefficient, describing the grams of predator produced per gram of prey consumed, and $w$ is the predation intensity.
Mirroring the calculation of the consumer yield coefficient, $Y_P = M_C E_C/B_{\lambda_P}$, where $E_C$ is the energy density of consumable biomass carried by herbivore prey, and the $B_{\lambda_P}$ is the lifetime energy requirement of the predator (app. C).
%($w=1$ denotes specialization, whereas $w<1$ denotes generalization). % of the predator on the consumer prey
% $C$, and $P$ is the density of the predator population.
% \begin{equation}
% \frac{{\rm d}}{{\rm d}t}C = \lambda^{\rm max}_{\rm C}RC- \left(\sigma \left(1 - \frac{R}{k}\right) + \mu + bP^* \right)C.
% \end{equation}
% The capture efficiency $b$ of the predator population describes the per-predator consumption rate of the herbivore prey $C$.

\begin{figure*}
    \centering
    \includegraphics[width=1\textwidth]{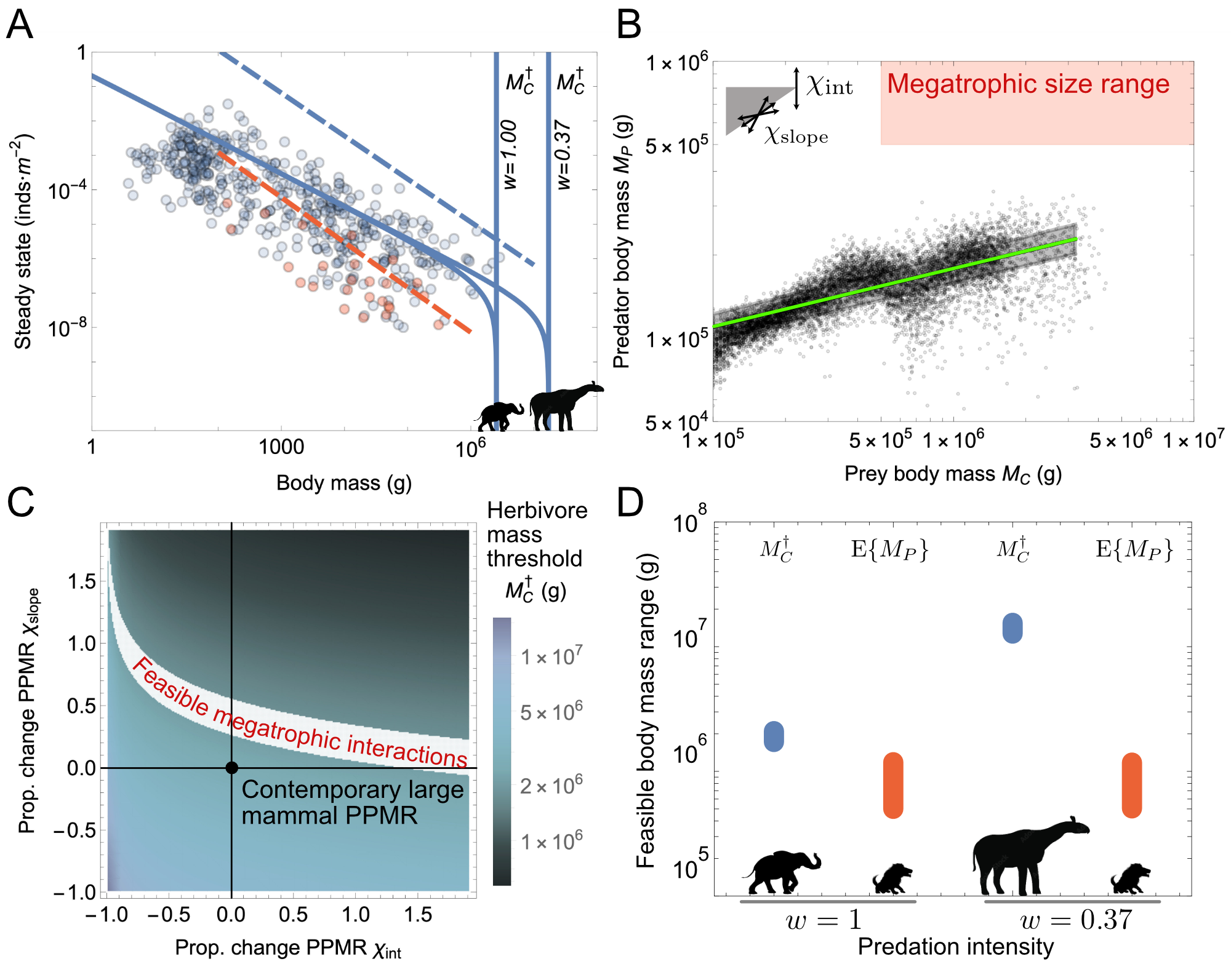}
    \caption{
          The feasibility of predation.
          A. Empirical mammalian herbivore (blue points) and predator (red points) mass-densities shown alongside the theoretical maximum herbivore $C^{\rm max}$ and predator $P^{\rm max}$ densities across body size (dashed lines). 
          The solid blue curves denote the predicted herbivore consumer steady state $C^*(M_C)$ with predation mortality given high ($w=1$) and low ($w=0.37$) predation intensities.
         % Predicted herbivore $M_C^\dagger$ and predator $M_P^\dagger$ size thresholds (blue and red vertical lines, respectively) resulting from a predation-induced instability.
         % (b) Percent mortality from predation for Serengeti herbivores (redrawn from Sinclair et al. \cite{Sinclair2003}) as a function of body mass $M_C$. Gray shaded region denotes a sigmoidal fit to the data with the inflection at $4.22\times10^5$ g. Vertical lines show herbivore size thresholds under the assumption of predator specialization ($f=1.00$; $M_C^\dagger=2.58\times10^6$) and generalization ($f=0.37$; $M_C^\dagger=1.75\times10^7$).
         B. Predator body mass as a function of prey body mass observed among contemporary mammalian fauna provides the allometric predator-prey mass relationship (PPMR).
         % Data bootstrapped from empirical observations in Refs. \citenum{hayward2005lion,Hayward2006hyena,hayward2006leopard,hayward2006lycaon,hayward2006cheetah,Hayward2008}.
         The green line denotes the best-fit, where $M_P = 9.76\times10^3 M_C^{0.21}$ g.
         % The red shaded region denotes megafaunal size ranges, which are not captured by contemporary predator-prey interactions.
         C. Threshold herbivore mass $M_C^\dagger$ given changes to the PPMR intercept $\chi_{\rm int}$ and slope $\chi_{\rm slope}$ (see Eq. \ref{eq:ppmr} and legend in B.).
         White shaded region highlights the mass range enabling feasible megatrophic interactions, shown in D. for herbivore mass thresholds $M_C^\dagger$ with the expected mass of associated carnivores ${\rm E}\{M_P\}$, assuming alternative predation intensities.
         % (d) As in (c), but showing the corresponding threshold carnivore sizes $M_P^\dagger$.
    }
    \label{fig:predrate}
  \end{figure*}

Assuming a linear functional response for predation mortality, $\lambda_P(C)$ is maximized when the consumer reaches its theoretical maximum population density, which we calculate by converting the resource carrying capacity directly to grams of consumer produced, or $C^{\rm max} = Y_C k$.
While this is an ultimately unattainable theoretical bound, it allows for a direct calculation of the predator growth rate as a function of $C$, written as
\begin{equation}
    \lambda_P(C) = \lambda_P^{\rm max}\frac{C}{C^{\rm max}} = \lambda_P^{\rm max}\frac{C}{Y_C k},
    \label{eq:predfunction}
\end{equation}
where $\lambda_P^{\rm max} = \ln(\nu)/t_{\lambda_P}$ is the maximum predator growth rate, given $\nu=2$, and $t_{\lambda_P}$ is the time required for the predator to reach maturity (following eq. \ref{eq:timescale}).
% assuming mammalian carnivore-specific metabolic relationships (see methods). 
% We assume that the predator capture efficiency is proportional to predator reproduction $\lambda^{\rm max}_P=\lambda_P/\hat{k}$, where the intrinsic reproductive rate $\lambda_P$ scales with maximal resource growth at one half resource carrying capacity $\hat{k}$.
% The capture efficiency can then be expressed as
% \begin{equation}
%     b = \frac{\lambda^{\rm max}_P}{Y_P Y_C},
%     % = \frac{\lambda_P}{Y_P(Y_C \hat{k})},
% \end{equation}
% where $Y_P$ is the predator yield coefficient describing the grams of predator produced per grams of herbivore prey consumed.
% Together, $Y_P Y_C$ thus represent the grams of predator produced per gram of resource consumed by its prey, mirroring our parameterization of the resource consumption rate in Eq. \ref{eq:2d}.
% Both the maximal rates of consumer growth $\lambda^{\rm max}_C$ and predator growth $\lambda^{\rm max}_P$ can be used to accurately predict the upper bounds of herbivore and carnivore population densities, respectively (see Supplementary Materials app. XX), lending particular support for our scaling of maximal consumer and predator growth rates. 
% Finally, because we do not explicitely include the full dynamics of the predator population, we assume predators exist at empirically measured steady state densities, where that $P\equiv P^*\propto M_P^{-0.88}$ \cite{carbone2002common}.  
The theoretical boundary density for herbivore consumers $C^{\rm max}$ can similarly be used to calculate the boundary density for predators, $P^{\rm max} = Y_P C^{\rm max}$, both of which accurately capture the upper-bounds of herbivore and carnivore mass-density observations (dashed lines in fig. \ref{fig:predrate}A).
Because the effects of the predator are implicit, we assume that the predator population remains at empirically measured steady state densities for mammalian carnivores, where $P^*= p_0 M_P^{p_1}$ given $p_0=8.62\times10^{-4}~{\rm inds}^{1-p_1}/{\rm m^2}$ and $p_1 = -0.88$ \cite{carbone2002common}.  
As we are employing this framework to evaluate longer-term evolutionary consequences, this condition assumes that predator densities do not have long-term feasibility if they stray far from $P^*$.

% %Predation less than max
% Though we are not explicitly modeling the predator population, we expect the per-predator capture efficiency to decline as the consumer population is below a critical density.

% [We've thus scaled the intrinsic predator growth rate in the same way as consumer growth... to the maximal growth of resources]
% [This represents maximum predator growth]
% [We can check that this scaling makes sense -- predicts upper boundary of predator data]

% [We observe that $\lambda_P^{\rm max}$ is the reproductive rate of the predator population per gram of herbivore prey consumed.]
% In this case we define this critical point by the half-saturation density of resources $\hat{k}$, such that $\lambda_P^{\rm max} = \lambda_P/\hat{k}$ where $\lambda_P$ is the predator's intrinsic rate of increase, and $Y_C \hat{k}$ represents the grams of consumer produced under the conditions of maximum resource growth.
% % , such that $b = \lambda_P^{\rm max}/Y_P$. % , such that $\lambda_P^{\rm max} = \lambda_P/\hat{k}_C$
% % The predator yield coefficient scales proportionally with the individual mass of the predator and the energy density of prey, normalized to the lifetime energetic needs of a predator individual reaching maturity.

The predation mortality rate depends on both the body size of the herbivore consumer and its respective predator. 
% The predator yield coefficient scales proportionately with predator body size and the energy density of its prey $E_C$, such that $Y_P \propto M_P E_C$.
% Because we are assuming that the predator population is external to our consumer-resource model, the mortality rate experienced by the herbivore population $bP$ is sensitive to predator size in two ways: \emph{i}) larger predator body sizes will increase the predator yield, lowering the attack efficiency
Trophic interactions are constrained by body size \cite{Sinclair2003,Brose2005,Hatton:2015fk}, though the nature of the predator-prey mass relationship (PPMR) varies across communities \cite{barnes2010global} and size classes \cite{Carbone:1999ju,Brose2005,Rohr2010,riede2011stepping,pires2015pleistocene,nakazawa2017individual}.
Compellingly, PPMRs for many clades can be predicted from the scaling of handling time \cite{delong2020}, suggesting that the signatures of body size evolution has cascading effects on community structure and function.
% outside of aquatic gape-limited systems \cite{Carbone:1999ju,nakazawa2017individual}. %and larger prey generally suffer mortality from proportionately smaller predators, 
While prior work has largely focused on the expected prey mass for a given predator mass, because our framework is prey-centric we require a prediction of the expected predator mass $M_P$ given an herbivore of body size $M_C$.
% , which cannot be directly extrapolated from the expected prey mass for a given predator mass (see app. C).
For larger predators and prey ($>10^5$ g), the expected predator mass given a particular herbivore mass follows roughly ${\rm E}\{M_P\} = v_0 M_C^{v_1}$, where $v_0 = 9.76\times10^3$ g${}^{1-v_1}$ and $v_1 = 0.21$ \cite[fig. \ref{fig:predrate}B; see app. C;][]{hayward2005lion,Hayward2006hyena,hayward2006leopard,hayward2006lycaon,hayward2006cheetah,Hayward2008}. %A) Hayward replicated data; B) pred size per herb size
%  9.76\times10^3 M_C^{0.21}
\cite[Here and throughout the prefix `mega' is used to signify size classes $>5\times10^5$ g;][]{hayward2005lion}.
Accordingly, larger terrestrial herbivores tend to suffer mortality from proportionately smaller predators, an asymmetry that becomes more pronounced with increasing size \cite[cf. ][]{Sinclair2003}.
We note that smaller terrestrial predator/prey size classes tend to be much larger than prey \cite[e.g. rodent- or insect-specialist mesocarnivores;][]{cruz2022geography,Cruz2022}, also captured by the ${\rm E}\{M_P\}$ scaling.
% , so we limit our current analysis to those occupying size classes $\geq 100$ Kg.

Integrating the large-bodied PPMR into the predation mortality rate reveals the emergence of a dynamic instability at megaherbivore size classes (fig. \ref{fig:predrate}A,B), the product of a transcritical bifurcation at consumer mass $M_C^\dagger$ (app. C), and similar to the more general instability documented by \citet{weitz2006size}.
An implicit predator population with body size ${\rm E}\{M_P\}$ is thus able to withdraw sufficient biomass from an herbivore population -- without crashing the herbivore population -- below a threshold herbivore size of $M_C^\dagger = 2.58\times 10^6$ g (fig. \ref{fig:predrate}A).
Above this critical size threshold, the herbivore population has such low densities that it is unable to sustain a specialist predator species large enough to consume it, introducing a strong upper-bound to mammalian carnivore body size driven by a trophic cascade.
This boundary matches the herbivore maximum size limit observed in contemporary terrestrial systems \cite{Sinclair2003}, at roughly the size of an elephant (fig. \ref{fig:predrate}B; app. C), though the exact placement of $M_C^\dagger$ varies with the resource growth rate and carrying capacity.
While we have assumed values representative of grass resources, decreasing $\alpha$ and/or $k$ lowers the steady state mass-density intercept (eq. \ref{eq:steadystate}), setting the mass threshold at a lower body size (app. C).
This means that lower-productivity environments, or environments subject to large and long-term oscillations in productivity, may be expected to have more severe limitations on feasible megaherbivore sizes. 
And while we do not consider stochastic or transient effects directly, we may also assume actualized extinction risk to emerge at smaller-than-predicted masses where transient or stochastic effects may push populations below the point of recovery.
% While the largest herbivores do not suffer significant mortality from predation in contemporary ecosystems, this boundary is critical because it signifies the threshold where excess mortality renders the population highly unstable.
% The herbivore mass threshold is not incredibly sensitive to variation, where a $\pm 10\%$ change in mortality rates results in a $M_C^\dagger$ ranging from $2.15$ to $3.16\times10^6$ g (Supplementary fig. XX), well within the elephant body mass range \cite{christiansen2004body}. %CHECK

$M_C^\dagger$ marks the threshold herbivore mass above which predation is unsustainable, though \citet{Sinclair2003} revealed contemporary herbivores to escape predation at ca. $4.22\times10^5$ g.
This change-point reflects the limitations of contemporary carnivores, which reach a maximum body size of $1.15$ to $2.60\times10^5$ g \cite{Sinclair2003}, and have preferences for prey up to $5.50\times10^5$ g \cite{hayward2005lion}.
Importantly, the sole predators of contemporary giants are not megaherbivore specialists, instead opportunistically subsidizing their preferred prey with larger taxa. %, while relying on a much smaller prey base.
% This highlights an important constraint of our framework: that predation suffered by the herbivore is exclusive to a single predator population, and that the single predator population is relying exclusively on a single prey.
While we have so far assumed a predator-prey interaction where the entirety of predator growth is fueled by the focal herbivore, the largest predators in natural systems tend be dietary generalists \cite{Sinclair2003,gross2009generalized}.
% In diverse mammalian communities, the largest predators also exhibit the largest prey size range \cite{Sinclair2003}, suggesting that only a fraction of predator biomass is expected to be supported by a single prey population.
% We can introduce the effects of generalist predation by modifying the fraction of predator growth $f$ supported by the herbivore prey, such that the demands of predator growth are modified as $f\lambda_P(C)$.
% to a fraction of its current estimate, effectively simulating the influence of a generalist predator population.
We observe that reducing the predation intensity (such that $w<1$) increases $M_C^\dagger$ to a larger threshold mass (app. C).
For example, $w=0.37$ increases the herbivore body mass boundary to $M_C^\dagger = 1.75\times10^7$ g (fig. \ref{fig:predrate}B; app. C), roughly the body mass attained by the largest terrestrial herbivores, the Oligocene paraceratheres and Miocene deinotheres \cite{Smith:2010p3442}. %means that a predator is supporting a little more than 1/3 of its growth rate by the targeted prey, increasing

That the threshold herbivore mass decreases with increasing predation intensity suggests that larger predators are dynamically constrained to be dietary generalists \cite{Sinclair2003}, while also pointing to an amplifying feedback mechanism \cite{brook2008synergies} that may operate in diverse communities undergoing megafaunal extinctions.
As megaherbivore species are lost, the largest predators must respond by increasing the intensity of predation on those remaining.
Our results suggest that this energetic redirection reduces the threshold herbivore mass $M_C^\dagger$ to lower size classes, increasing the likelihood of additional extinctions and attendant increases in predation intensity on survivors. 
Together, this demonstrates a dynamic mechanism for the previously proposed influence of top-down dietary ratcheting hypothesized for the Pleistocene extinctions, in particular \cite{kay2002false,ripple2010linking}. % and attendant predator mortality

%NOTE: THE THRESHOLD IS ALWAYS AT LARGER SIZES

%RELAXING THE PPMR
% Contemporary megaherbivores do not represent the largest mammalian size classes ($>3\times10^6$ to $1.74\times10^7$ g), and there are no contemporary megapredators.
Adaptations to hypercarnivorous strategies are difficult to reverse on macroevolutionary time scales, resulting in the so-called `hypercarnivore ratchet' \cite{VanValkenburgh1991,Holliday2004}.
Moreover, there is a strong correlation between hypercarnivorous adaptations and body size among terrestrial carnivores, the combination of which may promote vulnerability to extinction \cite{VanValkenburgh:2004p2451}.
While deinotheres and paraceratheres top the megaherbivore scale, the Eocene artiodactyl \emph{Andrewsarchus} may have been the largest terrestrial mammalian predator at ca. $1\times10^6$ g \cite{burness2001dinosaurs}, while the Miocene Hyaenodontid \emph{Megistotherium osteothlastes} ranged between $5$ to $8\times10^5$ g and the early Eocene Oxyaenodont \emph{Sarkastodon mongoliensis} weighed ca. $8\times10^5$ g \cite{sorkin2008biomechanical}. %(Rasmussin 1989, Sorkin 2008)
% a recent skull of \emph{Smilodon populator} indicates that some individuals may have attained body sizes up to 900 Kg (REF).
A theoretical maximum mammalian carnivore size of $1.1\times10^6$ g has been proposed based on the intersection of daily energetic uptake requirements against metabolic expenditures \cite{Carbone:2007dz}, closely aligning with the largest known megapredators.
While our consumer-resource framework provides a range of predicted megaherbivore body mass thresholds depending on the fraction of predator growth it fuels, we next ask under what conditions megatrophic relationships between megaherbivores and megapredators are dynamically feasible.

A central relationship in our framework is the allometric PPMR observed for the largest contemporary herbivores and carnivores, however empirical observations of PPMRs reveal tremendous variability both across and within clades \cite{delong2020}, and are completely unknown for megatrophic interactions in terrestrial systems (fig. \ref{fig:predrate}B; app. C). 
% this PPMR cannot account for past megatrophic relationships (Fi. \ref{fig:predrate}b; app. C).
% In fact, it is likely that the contemporary PPMR does not account for these interactions, as deinotherium-sized herbivores result in ${\rm E}(M_P|M_C)=2.99\times10^5$ g, whereas the maximum carnivore body mass is $2-3\times$ larger.
While it is unknown whether these super-sized carnivores were specialists on deinothere size-classes, our framework allows us to investigate whether and to what extent changes to the contemporary PPMR enable megatrophic interactions (fig. \ref{fig:predrate}C,D).
To examine this, we allow the expected predator mass given a particular prey mass to vary as 
\begin{equation}
{\rm E}\{M_P\} = v_0(1+\chi_{\rm int}) M_C^{v_1(1+\chi_{\rm slope})},
\label{eq:ppmr}
\end{equation}
where the proportional changes in the PPMR intercept and slope are given by $\chi_{\rm int}$ and $\chi_{\rm slope} \in (-0.99,2)$ (see the legend in fig. \ref{fig:predrate}B).
We note that while $\chi_{\rm slope}$ explores changes to the inferred steepness of the PPMR, $\chi_{\rm int}$ explores variation around the scaling relationship for a particular prey mass. 
% We find that the threshold herbivore body size $M_C^\dagger$ increases, while $M_P^\dagger$ decreases, with lower PPMR intercepts and shallower slopes (fig. \ref{fig:predrate}c,d). %, whereas the threshold carnivore body size $M_P^\dagger$ decreases with the same 
% Lower PPMR intercepts and shallower slopes mean that predator sizes are generally smaller, and increase more slowly, with larger herbivore body sizes.
% So given a particular herbivore size, proportionately smaller predators elevate the threshold herbivore mass, while proportionately larger predators drive down the threshold herbivore mass.
% % whereas smaller-bodied predators serve to elevate $M_C^\dagger$.
% Importantly, 

We observe that only a small range of values for PPMR intercepts and slopes permit the existence of dynamically feasible megatrophic interactions, where megaherbivores serve as prey for specialized megapredators (white band in fig. \ref{fig:predrate}C,D; app. C). %($1.72-2.01\times10^6$ g) ($6.11\times10^5-1.19\times10^6$ g;
% While significantly larger PPMR intercepts ($\chi_{\rm int}\gg 0$) are unlikely to be realized in natural systems, the megainteraction range does include very low intercepts with very high slopes, such that predator mass increases steeply with increasing prey mass.
Such interactions could be realized if the PPMR in fig. \ref{fig:predrate}B had an increased intercept or alternatively both a lower intercept and higher slope for mega size-classes.
That alternative PPMRs could characterize different size-classes across foraging guilds has been previously examined \cite{vezina1985empirical,Carbone:1999ju}, and the clear disconnect between that shown for contemporary large-bodied mammals and those in the megatrophic range (fig. \ref{fig:predrate}B) supports this notion.
When predation intensity is high ($w=1$), the PPMR enabling feasible megatrophic interactions lowers the herbivore mass threshold $M_C^\dagger$, resulting in megapredators consuming relatively smaller prey.
However, if predation intensity is lowered ($w=0.37$), we observe feasible megatrophic interactions for size-classes capturing megaherbivores and megapredators at their largest documented sizes in the fossil record (fig. \ref{fig:predrate}D; app. C).
% Incorporating a PPMR that enables feasible megatrophic interactions, while permitting the existence of larger predators, does so by lowering the herbivore mass threshold $M_C^\dagger$ (fig. \ref{fig:predrate}C).
% Because we have so far considered only high intensity predation $(w=1)$,
% such that PPMRs are low at smaller masses, and much higher at large masses.
% Such a hypothesized PPMR does not stray far from contemporary large-mammal interactions  (app. C) and may be a good candidate for megatrophic interactions. 
% Feasible megatrophic interactions increase substantially if a smaller percentage of the predator growth rate is fueled by the target herbivore population ($w<1$).
% Setting $w=0.37$ -- which we observed increases $M_C^\dagger$ to deinothere/indricothere size classes  -- and allowing both $\chi_{\rm int}$ and $\chi_{\rm slope}$ to vary, results in megatrophic interactions spanning both the largest megaherbivore and megapredator sizes observed in the fossil record (app. C). %($6.15\times10^5-1.20\times10^6$ g) ($1.27-1.48\times10^7$ g)
% These results indicate that existence of megapredators in terrestrial systems requires generalist dietary behaviors, and only steeper PPMRs than those observed in contemporary systems enable interactions of the largest mammalian size classes.
% However, as ever-increasing predator body sizes may expand the size range of available herbivore prey, the increased demands of these predator populations lower the maximum sustainable herbivore body size.
% The sensitivity of this instability suggests that 

That decreased predation intensity enables feasibility of the largest mammalian size-classes agrees with previous conjectures that the largest mammalian terrestrial predators were likely dietary generalists \cite{farlow1993rareness}.
Our framework thus highlights dynamic constraints existing between predators and prey that may serve to structure mammalian communities over evolutionary time, in particular revealing the susceptibility of megaherbivores to perturbations.
As carnivorous clades evolved body sizes enabling megaherbivore predation, their super-sized appetites may have suppressed megaherbivores to unsustainable densities where the risk of extinction became overwhelming -- an evolutionary trap marking the final tooth in the hypercarnivore ratchet \cite[cf.][]{VanValkenburgh:2004p2451}.

\subsection*{Harvesting to extinction} 
% (extrinsic, mass-non-specific, large-size punishing)}

We last consider the effects of anthropogenic harvest-induced mortality on herbivore populations.
While the predation rate is naturally limited by the energetic needs of the predator, we consider harvest to be a comparatively unconstrained source of mortality.
This may be the case if the human population(s) engaged in harvesting are subsidized by alternative resources \cite{Brook2005}.
Harvest pressure has potentially varying relationships with consumer (prey) body mass, a complex product of environment, climate, culture, and technology \cite{churchill1993weapon}.
% , where we can assume a general harvest rate is of the form $h = \xi M_C^w$.
For example, hunting traditions specializing in mass-collecting, by way of trapping or netting \cite{churchill1993weapon,ugan2005does} are expected to exhibit harvest allometries biased towards smaller species, whereas a purely opportunistic strategy may be expected to have very little allometric dependence.
 % (negative size-scaling, $w<0$) (zero size-scaling, $w=0$).
% Inclusion of negative size-scaling harvest reveals that smaller-sized prey can withstand significant harvesting pressure before their populations are negatively impacted.
While smaller mammals do not appear to offer a significant return on investment, the mass-collecting of invertebrates, such as grasshoppers, and fish can offer significant returns \cite{ugan2005does}.
% The inclusion of zero size-scaling harvest reveals that it is the larger-bodied species that are negatively impacted, but only where $h(M_C) > \lambda_C(M_C)$ (app. D).
In contrast, the innovation of advanced projectiles is thought to have enabled harvest of terrestrial megafauna \cite{churchill1993weapon,prates2022changes}, and archeological evidence points to many Pleistocene human populations as potential megafaunal specialists \cite{Smith:2018gm}. %(positive size-scaling, $w > 0$)

% Technology provides access to otherwise unobtainable size-classes of potential prey, and is expected to strongly influence the allometry of harvesting effort.
% For example, the innovation of advanced projectiles is thought to have enabled harvest of terrestrial megafauna \cite{churchill1993weapon,prates2022changes}, while archeological evidence points to many Pleistocene human populations as potential megafaunal specialists (positive size-scaling, $w > 0$) \cite{Smith:2018gm}.
% More recently, historical fisheries have been subject to positive size-scaling harvest efforts \cite{Sethi2010}, and contemporary size classes may be much reduced due to the accumulation of these pressures over time \cite{jennings2004fish}.

\begin{figure*}
    \centering
    \includegraphics[width=1\linewidth]{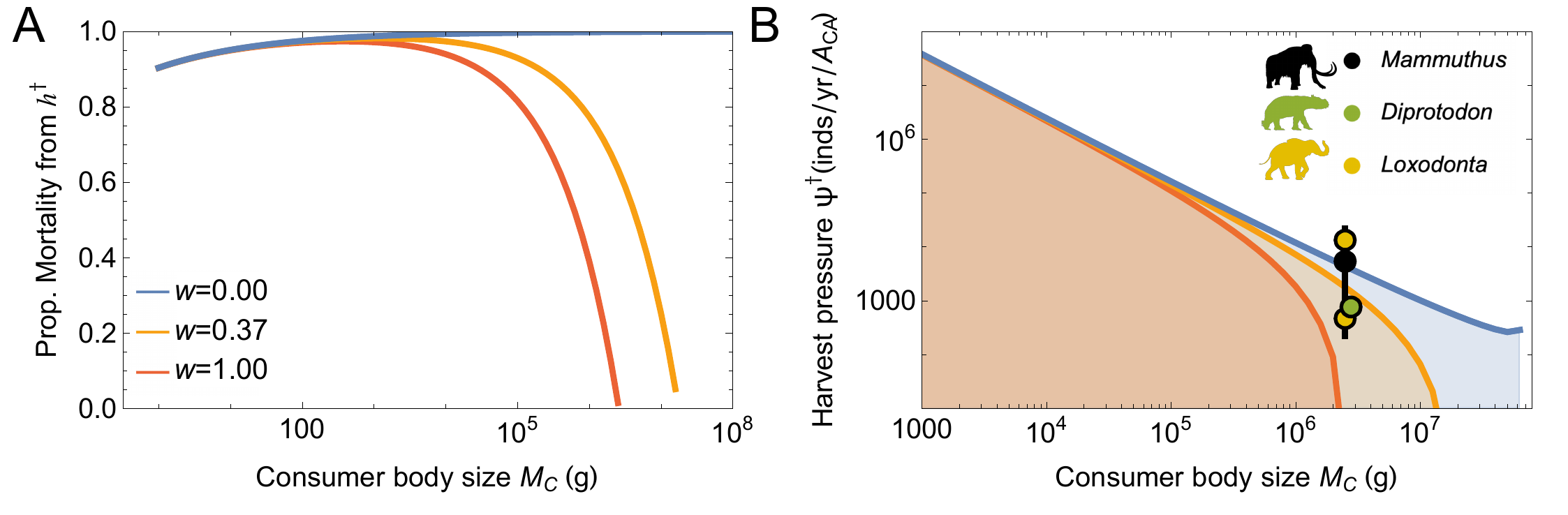}
    \caption{
        The effects of harvest mortality on herbivore consumers.
        A. Proportion mortality due to an extinction-inducing harvest rate $h^\dagger$ without predation ($w=0$; blue line), and with low ($w=0.37$; orange line) or high predation intensity ($w=1$; red line), as a function of consumer body mass $M_C$.
        B. Harvest pressure $\psi^\dagger$ resulting from extinction-inducing harvest (inds/year/$A_{\rm CA}$) without predation ($w=0$; blue line), and with low ($w=0.37$; orange line) or high predation intensity ($w=1$; red line), as a function of consumer body mass $M_C$.
        Black point and line: median and range of estimated harvest rates for woolly mammoths \cite[\emph{Mammuthus primigenius};][]{fordham2022process}; Green point: estimated harvest pressure for the Australian \emph{Diprotodon} \cite{bradshaw2021relative}; lower and higher yellow point: estimated harvest rates for contemporary \emph{Loxodonta} during the early 1800s and just prior to 1987, respectively \cite{milner1993exploitation}.
    }
    \label{fig:harvest}
\end{figure*}

Because harvest scaling may be difficult to measure and idiosyncratic, we instead calculate the harvest rate required to induce extinction, $h^\dagger$, as a function of body size $M_C$, and find a scaling relationship proportional to the rate of reproduction where $h^\dagger \propto M_C^{-1/4}$.
% We denote the harvest rate sufficient 
This is a natural result, as the effort required to suppress a population is expected to be proportional to its reproductive rate, reflecting the increased susceptibility of large-bodied organisms to extinction \cite{enquist2020megabiota}. 
As a proportion of the other sources of consumer mortality that we have considered (excluding predation; $w=0$), extinction-level harvesting is lower for smaller consumers, saturating at close to unity at large size classes, reflecting the elevated role of starvation mortality among smaller-sized organisms (Fig \ref{fig:harvest}A).
With predation mortality included at both low ($w = 0.37$) and high ($w=1$) intensities, extinction-level harvesting accounts for an increasingly smaller proportion of mortality for larger organisms (orange and red lines, fig. \ref{fig:harvest}A).
This highlights the delicate nature of the megafaunal niche, where smaller changes in mortality rates can induce population collapse \cite{enquist2020megabiota}.

To examine how our estimate of extinction-level harvesting rates $h^\dagger$ compare to those estimated for human hunting of paleontological and historical mammalian populations, we converted $h^\dagger$ to harvest pressure $\psi^\dagger$, or the number of individuals harvested per year to reduce the population to a fraction of its steady state $\epsilon C^*$ where we set $\epsilon = 0.01$.
We calculate $\psi^\dagger$ for an arbitrary area (see app. D), which we standardize to the area of California ($A_{\rm CA} = 4.24\times10^5~{\rm km}^2$), such that
\begin{equation}
    \psi^\dagger \propto -h^\dagger\frac{C^*(1-\epsilon)}{M_C \log(\epsilon)}.
\end{equation}
Though the annual harvesting pressure is unrealistically high for smaller organisms, we observe that it is ca. $4.3\times10^3~{\rm inds/yr/}A_{\rm CA}$ for elephant-sized mammals (ca. $2.5\times10^6$ g) in the absence of predation mortality ($w=0$).
With increasing predation intensity, the harvest pressure required to induce extinction is much less for these larger consumers (orange and red lines in fig. \ref{fig:harvest}B).
We note that this calculation of harvest pressure should be viewed as a minimum estimate given that we do not account for demographic rebound.
As such, this measure is appropriate only if the timescale of harvest is less than the generational timescale, which is the case for the megafauna considered here. %is derived from an extinction timescale that 

% , and $>1.8\times10^6$ g in the presence of predation mortality (fig. \ref{fig:harvest}B).

% Our estimate of harvest pressure, while assuming a constant depletion from the population steady state such that it is conservative on greater-than-generational timescales (see Supplementary Materials XX), also permits direct comparison to other published estimates.
Our predictions of extinction-inducing harvest pressure compare well with paleontological and historical estimates of harvest pressure on mammalian megafauna (fig. \ref{fig:harvest}B; see app. D). 
For example, using a formulation similar to that of \citet{alroy2001multispecies}, \citet{fordham2022process} estimate the harvest pressure required to collapse mammoth (\emph{Mammuthus primigenius}) populations, revealing a range of values consistent with our expectation for similar size-classes (est. $\psi^\dagger =$ ca. $1.24\times10^4~{\rm inds/yr/}A_{\rm CA}$), as did estimates of extinction-inducing harvest of the Australian \emph{Diprotodon} \cite[est. $\psi^\dagger =$ ca. 763 ${\rm inds/yr/}A_{\rm CA}$;][]{bradshaw2021relative}.
Within the historical record, elephant (\emph{Loxodonta}) populations experienced comparatively lower harvest pressure through 1850 \cite[ca. $466~{\rm inds/yr/}A_{\rm CA}$, derived from the volume of ivory exports;][]{milner1993exploitation}.
% , scaled to African elephant habitat area ($3.22\times 10^{6}~{\rm km}^2$; Thouless, 2016).
While fluctuating over the next century, harvest pressure increased to a maximum of ca. $13.3\times 10^5~{\rm inds/yr/}A_{\rm CA}$ just prior to 1987 (fig. \ref{fig:harvest}B).
This level of harvest was not sustained, as ivory export volume plummeted following the implementation of trade restrictions in 1989 \cite{milner1993exploitation}.
% While this level of harvest is greater than our estimate to induce population collapse (fig. \ref{fig:harvest}B), it was not sustained, as ivory export volume plummeted following the implementation of trade restrictions in 1989 (Milner-Gulland 1993).
Both the \citet{fordham2022process} estimate for Pleistocene mammoths and the short-lived harvest maximum for African elephants in 1987 \cite{milner1993exploitation} achieved pressures greater than $\psi^\dagger$ under the conservative assumption of no natural predation (fig. \ref{fig:harvest}B).
While estimates for $\emph{Diprotodon}$ harvest are considerably lower \cite{bradshaw2021relative}, it is important to note that our framework is parameterized for eutherian rather than marsupial mammals.
Nevertheless, the estimated \emph{Diprotodon} $\psi^\dagger$ is well within range of extinction-inducing harvest rates if natural predation pressures are also included, and there is evidence to suggest that \emph{Diprotodon} likely served as prey for marsupial lions \cite{horton1981cuts,wroe1999estimating}, and both giant crocodylians (\emph{Pallimnarchus} spp.) and varanid lizards \cite[\emph{Megalania} spp.;][]{webb2009late}.
% (Runnegar 1981; Horton 1981; Wroe et al. 1999) 
% though these estimates must be viewed as lower-bounds given they are derived from ivory exports extracted from an unknown area, though standardized Africa, much larger than the area being harvested.

\section*{Conclusion}

We have shown that the inclusion of mass-specific energetic transfer between resources and consumers, combined with the unique timescales governing consumer mortality, both predict Damuth's Law \cite{Damuth1987} and provide insight into dynamic thresholds constraining populations.
While natural and starvation mortality primarily impact small-bodied species, trophic mortality primarily impacts large-bodied species with longer generational timescales. 
Moreover, while mass-specific predation gives rise to dynamic thresholds for herbivore populations, these effects are sensitive to both predation intesnity as well as the associated predator-prey mass relationship, which isn't well understood in terrestrial ecosystems \cite{nakazawa2017individual}.
% We suggest that our minimal energetic framework both captures the largest-scale constraints shaping mammalian communities, while simultaneously underscoring the fragility mammalian megafauna.
While assessment of particular communities and/or species requires more detailed approaches -- integrating, for example, life history dynamics as in \citet{bradshaw2021relative} -- we suggest that a lower-dimensional framework is useful for extracting general, first-order energetic constraints that both shape and potentially limit the nature of mammalian communities.

That extinction risk appears to increase with body size \cite{Cardillo:2005et} is integral to our understanding of the Pleistocene extinctions \cite{alroy2001multispecies,johnson2002determinants,brook2008synergies,Smith:2018gm,bradshaw2021relative} and anthropogenic effects throughout the Holocene \cite{Estes:2011eo}.
% That megafauna are ephemeral and more susceptible to external sources of mortality is a cornerstone of Quaternary paleontology \cite{Brook2005}.
Because megafaunal loss may have disproportionately large impacts on ecosystem functioning \cite{enquist2020megabiota}, understanding the mechanistic drivers that may lead these species to the brink is of paramount importance.
Assessing which energetic walls close in and why as body size increases, is a fundamental aspect of reconciling the nature of extinction \cite{brook2008synergies}, particularly when there is size-selectivity \cite{Smith:2018gm}.
% Because energetic compartment models broadly apply across clades, they are particularly useful for exploring the constraints of paleocommunities.
That we observe dynamically-feasible megatrophic interactions to occupy a narrow band of predator-prey mass relationships points to a broader range of interaction structures than are realized in contemporary communities.
As the threshold consumer mass decreases with increased predation intensity, how megafaunal trophic structure changes during extinction cascades may be central for understanding the dynamics of community disassembly \cite{Yeakel2014}.
And while these dynamics may arise naturally from the energetic limitations of mammalian interactions, it may be that the added pressure of subsidized harvest, particularly on megafauna, inevitably leads to collapse.

% \clearpage

\begin{table*}[h]
\renewcommand{\arraystretch}{0.8}
\caption{Model parameters and values/units}
\label{tab:param}
    \begin{center}
    \footnotesize
         \begin{tabular}{lll}
            \hline
            Definition & Parameter & Value/Units   \\ %& References
            \hline
            Resource & & \\
            ~~~~~density & $R$ & ${\rm g/m^2}$ \\
            ~~~~~reproduction rate & $\alpha$ & $9.49\times10^{-9}~({\rm 1/s})$ \\
            ~~~~~carrying capacity & $k$ & $23\times10^3~({\rm g/m^2})$ \\
            ~~~~~energy density & $E_d$ & $1.82\times 10^4$~(J/g) \\
            \hline
            Consumer & & \\
            ~~~~~density & $C$ & ${\rm g/m^2}$ \\
            ~~~~~body mass & $M_C$ & g \\
            % ~~~~~body mass threshold & $M_C^\dagger$ & g \\
            % ~~~~~theoretical max & $C^{\rm max}$ & ${\rm g/m^2}$ \\
            ~~~~~timescale of growth from $m_1$ to $m_2$ & $\tau(m_1,m_2)$ & s \\
            ~~~~~reproduction rate & $\lambda_C^{\rm max}$ & ${\rm 1/s}$ \\
            ~~~~~yield coefficient & $Y_C$ & $({\rm g/m^2~}C)/({\rm  g/m^2~} R)$ \\
            ~~~~~maintenance rate & $\rho$ & ${\rm 1/s}$ \\
            ~~~~~natural mortality rate & $\mu$ & ${\rm 1/s}$ \\
            ~~~~~starvation rate & $\sigma^{\rm max}$ & ${\rm 1/s}$ \\
            ~~~~~harvest rate & $h$ & ${\rm 1/s}$ \\
            \hline
            Predator & & \\
            % ~~~~~steady state intercept${}^1$ & $P_0$ & $8.62\times10^{-4}~{\rm inds/m^2}$ \\
            ~~~~~steady state density${}^1$ & $P^*$ &  $P_0 M_P^{-0.88}~{\rm inds/m^2}$ \\
            ~~~~~body mass & $M_P$ & g \\
            % ~~~~~body mass theshold & $M_P^\dagger$ & g \\
            % ~~~~~theoretical max & $P^{\rm max}$ & ${\rm g/m^2}$ \\
            ~~~~~growth rate & $\lambda_P^{\rm max}$ & ${\rm 1/s}$ \\
            ~~~~~yield coefficient & $Y_P$ & $({\rm g/m^2~}P)/({\rm  g/m^2~} C)$ \\
            ~~~~~predation intensity & $w$ & (0,1) \\
            \hline
            % Prop. change starvation rate & $\chi_s$ & (-0.99,1) \\
            % PPMR intercept${}^2$ & $v_0$ & $1.18\times10^5$ g \\
            % PPMR slope${}^2$ & $v_1$ & $0.19$ \\
            Metabolic normalization constant & $B_0$ & 0.047 (W g${}^{-3/4}$) \\
            Energy to synthesize a unit of mass & $E_m$ & 5774 (J g${}^{-1}$) \\
            Energy stored in a unit of mass & $E_m^\prime$ & 7000 (J g${}^{-1}$) \\
            Prop. change PPMR intercept${}^1$ & $\chi_{\rm int}$ & (-0.99,2) \\
            Prop. change PPMR slope${}^1$ & $\chi_{\rm slope}$ &  (-0.99,2) \\
            Extinction-inducing harvest rate & $h^\dagger$ & ${\rm 1/s}$ \\
            Extinction-inducing harvest pressure & $\psi^\dagger$ & inds/yr/$A_{\rm CA}$ \\
            % Post-harvest fraction consumer density & $\epsilon$ & 0.01 \\
            \hline
            \multicolumn{2}{l}{\footnotesize{${}^1$PPMR: Predator-Prey Mass Relationship}}\\
        \end{tabular}
    \end{center}
\end{table*}

%%%%%%%%%%%%%%%%%%%%%
% Acknowledgments
%%%%%%%%%%%%%%%%%%%%%
% You may wish to remove the Acknowledgments section while your paper 
% is under review (unless you wish to waive your anonymity under
% double-blind review) if the Acknowledgments reveal your identity.
% If you remove this section, you will need to add it back in to your
% final files after acceptance.

\section*{Acknowledgments}

We would like to thank Irina Birskis-Barros, Uttam Bhat, Jessica Blois, Nathaniel Fox, Jacquelyn Gill, Paulo Guimar\~aes Jr., Emily Lindsey, Mathias Pires, Megha Suswaram, and Ritwika VPS for insightful comments and discussions that greatly improved the ideas and concepts that contributed to this manuscript. These ideas benefited greatly from travel funds provided to JDY from the Santa Fe Institute. This project was supported by National Science Foundation grant EAR-1623852 to JDY.

%%%%%%%%%%%%%%%%%%%%%
% Statement of Authorship
%%%%%%%%%%%%%%%%%%%%%
% This section should also be commented out while your MS is undergoing
% double-blind review. The specifics should of course be adapted to
% your paper, but the paragraph below gives some hints of possible
% contributions.

\section*{Statement of Authorship}

JDY, CPK, and TR conceived of the model.
JDY and TR developed the code and oversaw model analysis. 
All authors reviewed and edited the writing at all stages of composition.

\section*{Data and Code Availability}

Code and data archived on Zenodo: \texttt{https://doi.org/10.5281/zenodo.8213158}

% \defaultbibliographystyle{amnatnat}

% % \clearpage

% \defaultbibliography{aa_starving3}

% \putbib

\end{bibunit}

\clearpage

\renewcommand{\thesection}{Appendix \Alph{section}:}
\renewcommand{\thepage}{A\arabic{page}}
\renewcommand{\thetable}{A\arabic{table}}
\renewcommand{\thefigure}{\Alph{section}\arabic{figure}}
\renewcommand{\theequation}{A\arabic{equation}}
\renewcommand{\figurename}{Figure}
\setcounter{figure}{0}

\begin{bibunit}

\section{Natural mortality}

The natural mortality rate is obtained by first assuming that the number of surviving individuals in a cohort $N$ follows a Gompertz relationship \citep{CalderIII:1983jd}, where
\begin{equation}
    N = N_0 {\rm exp}\left( \frac{q_0}{q_a}\Big(1 - {\rm exp}({-q_a t})\Big)\right),
\end{equation}
given that $q_0$ is the initial cohort mortality rate, and $q_a$ is the annual rate of increase in mortality, or the actuarial mortality rate.
The change in the cohort's population over time then follows
\begin{equation}
    \frac{\rm d}{\rm dt}N = -d N,
\end{equation}
such that
\begin{equation}
    d = -\frac{1}{N}\frac{\rm d}{\rm dt}N.
\end{equation}
If $t_{\rm \ell}$ is the expected lifetime of the organism, then the average rate of mortality over a lifetime $t_\ell$ is
\begin{align}
    \mu &= \frac{1}{t_{\rm \ell}}\int_0^{t_{\rm \ell}}q_0{\rm exp}(q_a t_\ell) \nonumber \\ 
    &= \frac{q_0}{q_a t_\ell}\Big({\rm exp}(q_a t_\ell) - 1\Big).
\end{align}

The cohort mortality rate $q_0$, the actuarial mortality rate $q_a$ and the expected lifetime $t_\ell$ of a mammal with mass $M_C$ all follow allometric relationships, where $q_0 = 1.88\times 10^{-8} M_C^{-0.56}$ (1/s) and $q_a = 1.45\times 10^{-7} M_C^{-0.27}$ (1/s) where $M_C$ is in grams.
Together, we obtain the allometric relationship
\begin{equation}
    \mu(M_C) =\frac{3.21\times 10^{-8} \left({\rm exp}({0.586 M_C^{0.03}})-1\right)}{M_C^{0.59}}.
\end{equation}

\section{Variations in model parameters and allometric rates}
\setcounter{figure}{0}

While our framework dictates that plant growth rates and carrying capacities are directly proportional to consumer steady states, we can gain insight into what drives the very large range of observed consumer densities by exploring the observed ranges of $\alpha$ and $k$ in terrestrial systems.
We assume an intrinsic growth rate roughly that of grass where $\alpha = 9.45\times 10^{-9}$ (s${}^{-1}$), whereas observations among terrestrial plants reveal a range in growth rates from $2.81\times 10^{-10}$ to $2.19\times 10^{-8}$ \citep{michaletz2014convergence}, according with a change in $\alpha$ of roughly 97\% lower and 130\% higher than the set value. 
By incorporating this range into the estimated resource growth rate, we observe that we can account for a large portion of consumer steady state densities around the mean density (inner shaded region, Fig. 1, main text).
If we additionally adjust the carrying capacity $k$ of the resource to 90\% less-than and 150\% more-than the assumed value of $23\times 10^3$ g/m${}^2$, our framework accounts for nearly the full range of mammalian steady state densities (outer shaded region, Fig. 1, main text).
In this context, the upper-boundary of $k$ observed to capture most higher herbivore densities is ca. 34 kg/m${}^2$, which is on the higher end of estimated live above-ground biomass densities in terrestrial forests such as in Isle Royal and the Allegheny National Forest \citep{de2017simulating}.
% With an energy density of ca. 5.77 kiloJoules/gram, higher than but within range of the carrying capacity of $664\times10^3$ kJ/m${}^2$, [within the range of possibility for terrestrial systems, REF].

%Variations in steepness of slope
Our model's ability to capture the bounds of mammalian densities at low and high productivity invites some speculation into the actual steepness of the mass-density relationship.
While the best-fit slope to Damuth's Law is -0.77 we also observe that the steeper relationship given by our framework better captures the boundaries of mass-density data, whereas varying the intercept of the statistical best-fit would not capture the lower-density outer-boundary of larger species. %(-0.74 using XX) - especially those between $10^4$ and $10^5$ grams.
While within-clade mass-density relationships often reveal a shallower slope than if measured across clades \citep{Pedersen:2017he}, it is possible that the absence of data for larger mammals may bias estimates of the slope towards smaller (shallower) values.
Mammalian communities have undergone significant anthropogenic restructuring throughout the Holocene such that many larger species are excluded from the mass-density relationship by way of extinction \citep{Koch:2006vt}, and the greater prevalence of smaller species may introduce size-dependent biases.
For example, if species $<100$ g are excluded, the empirical mass-density slope steepens from $-0.77$ to $-0.85$.
% Because of the extinction-related bias of contemporary mammalian faunas (REF), the over-representation and influence of extant size classes should be considered when evaluating macroecological relationships (REF). 

\begin{figure}
  \centering
% \begin{subfigure}
\includegraphics[width=1\linewidth]{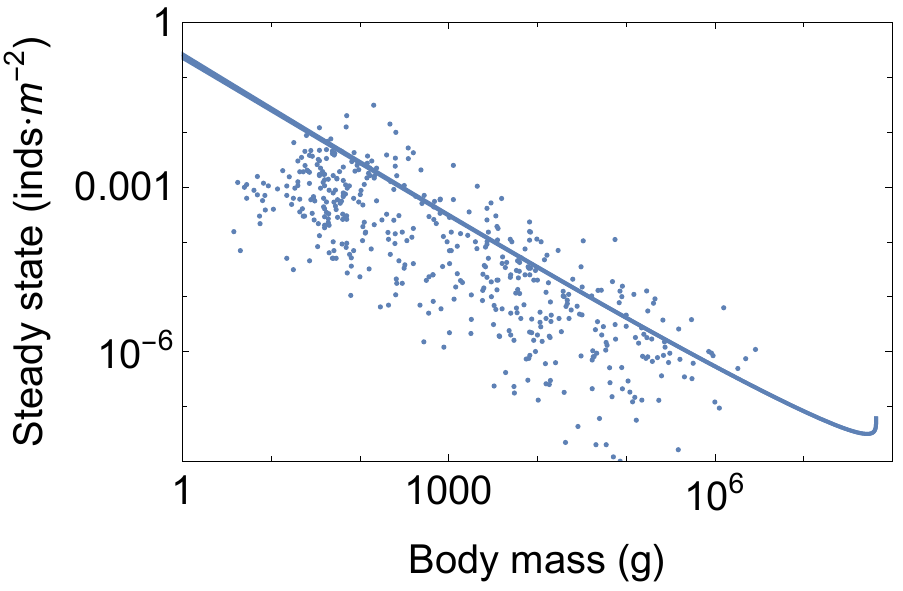}
% \end{subfigure}
% \begin{subfigure}
% \includegraphics[width=0.45\textwidth]{fig_slopemod.pdf}
% \includegraphics[width=0.9\textwidth]{starvation_reproductionrates.png}
% \end{subfigure}
\caption{
\footnotesize{
The predicted mass-density relationship with $\rho$ set to its allometric quantity versus $\rho = 0$. The two relationships cannot be distinguished, allowing us to extract insight from the consumer steady state (eq. 7, main text) without its inclusion.
}
}
\label{fig:rho}
\end{figure}

Considering how variations to the underlying energetic parameters driving consumer-resource dynamics alters the expected mass-density relationship may shed light on key constraints shaping mammalian communities. 
% PARAMETERS INFLUENCING INTERCEPT ONLY
We next explore how variations in the vital rates included in the consumer-resource model modify the expected intercept and slope of the mammalian mass-density relationship.
Different vital rates impact the mass-density relationship in three distinct ways, by either \emph{i}) influencing only the mass-density slope, \emph{ii}) influencing only the mass-density intercept, or \emph{iii}) influencing both.
Aside from the resource growth rate and carrying capacity, our framework also includes the intrinsic consumer reproductive rate $\lambda^{\rm max}_C$, the consumer yield coefficient $Y_C$, and the maximum rate of starvation $\sigma^{\rm max}$.
We introduce changes to these rates as, for example, $\lambda^{\rm max\prime}_C = \lambda^{\rm max}_C(1+\chi)$, where $\chi \in (-1,2)$ represents the proportion increase or decrease of the altered parameter denoted by ${}^\prime$.
We note that the recovery rate $\rho$ is sufficiently small that alterations do not have an influence on either the consumer mass-density intercept or slope (fig. \ref{fig:rho}).

Ignoring the effects of $\rho$, we can more easily intuit analytical expressions of the steady state conditions for both the consumer $C$ and resoure $R$, where
\begin{align}
C^* &= \alpha k Y_C \frac{\sigma^{\rm max}-2\lambda_C^{\rm max} + \sqrt{8{\sigma^{\rm max}}^2+(\sigma^{\rm max}-2\lambda_C^{\rm max})^2}}{4 {\sigma^{\rm max}}^2} \nonumber \\
R^* &= k \frac{\sigma^{\rm max} - 2\lambda_C^{\rm max}+ \sqrt{8{\sigma^{\rm max}}^2+(\sigma^{\rm max}-2\lambda_C^{\rm max})^2}}{4 \sigma^{\rm max}}.
\end{align}
We thus observe that the consumer steady state can also be expressed as
\begin{equation}
C^* = \frac{\alpha}{\sigma^{\rm max}}Y_C R^*.
\end{equation}
We discuss how the specific mass-scalings of the relationships impacting the steady states provide more intuition into Damuth's mass-density relationship in the main text.

\begin{figure}
  \centering
% \begin{subfigure}
\includegraphics[width=1\linewidth]{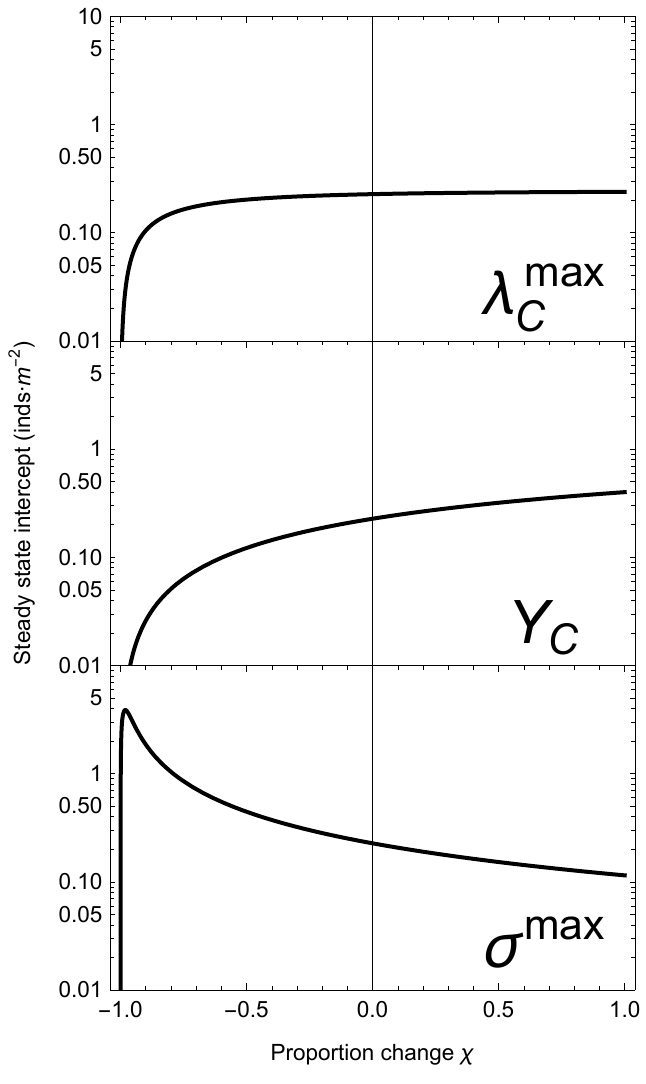}
% \end{subfigure}
% \begin{subfigure}
% \includegraphics[width=0.45\textwidth]{fig_slopemod.pdf}
% \includegraphics[width=0.9\textwidth]{starvation_reproductionrates.png}
% \end{subfigure}
\caption{
\footnotesize{
The effects of changes to metabolic parameters on the prediction of the mass-density relationship.
}
}
\label{fig:corr1}
\end{figure}

\begin{figure}
  \centering
  % \begin{subfigure}
  % \includegraphics[width=0.45\textwidth]{fig_intmod.pdf}
  % \end{subfigure}
  % \begin{subfigure}
  \includegraphics[width=1\linewidth]{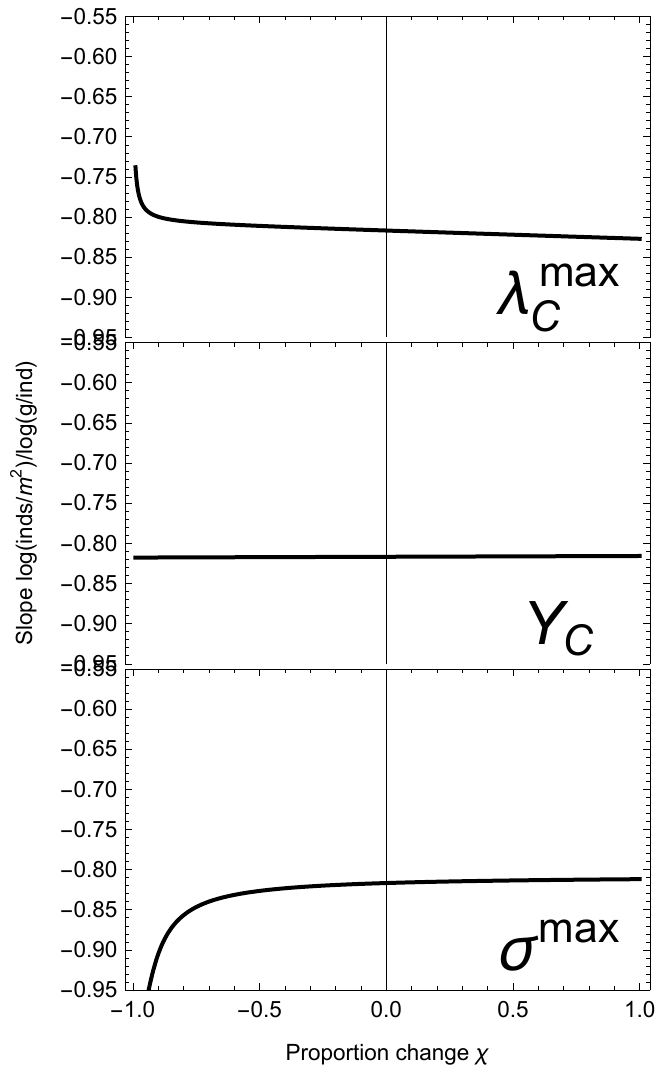}
  % \includegraphics[width=0.9\textwidth]{starvation_reproductionrates.png}
  % \end{subfigure}
  \caption{
  \footnotesize{
  The effects of changes to metabolic parameters on the prediction of the mass-density relationship.
  }
  }
  \label{fig:corr2}
\end{figure}

We can gain additional insight into the role of each vital rate by exploring their quantitative effects on the mass-density relationship directly.
Changes to the starvation rate have a large effect on both the consumer-density intercept and slope (Figs. \ref{fig:corr1},\ref{fig:corr2}).
We observe that decreasing $\sigma^{\rm max}$ from the expected value ($\chi<0$) serves to increase the steady state intercept, while decreasing the mass-density slope.
By comparison, increasing $\sigma$ from the expected value ($\chi>0$) has less effect on the mass-density relationship.
In the consumer-resource model described in Eq. 2.3 (main text), starvation is the primary source of consumer mortality, and therefore plays an out-sized role in determining consumer steady states. 
As this mortality is reduced, consumer densities increase, raising the intercept.
However, as consumer starvation rates decline we observe a steeper mass-density slope.
Reduced starvation rates therefore principally benefit the steady state densities of smaller species, with reduced effects observed for larger-bodied mammals.
Because fat biomass scales super-linearly with body mass (see Table 1, main text), the populations of larger consumers are more resilient to the effects of starvation, whereas those of smaller consumers are more prone.

% Changes to the starvation rate have a large effect on both the consumer-density intercept and slope (Fig. \ref{fig:corr}E,F).
% We observe that decreasing $\sigma$ from the expected value ($\chi<0$) serves to increase the steady state intercept, while decreasing the mass-density slope.
% By comparison, increasing $\sigma$ from the expected value ($\chi>0$) has less effect on the mass-density relationship.
% In the consumer-resource model described in Eq. \ref{eq:2d}, starvation is the sole source of consumer mortality, and therefore plays an out-sized role in determining consumer steady states. 
% As this mortality is reduced, consumer densities increase, raising the intercept.
% However, as consumer starvation rates decline we observe a steeper mass-density slope.
% Reduced starvation rates therefore principally benefit the steady state densities of smaller species, with reduced effects observed for larger-bodied mammals.
% Because fat biomass scales super-linearly with body mass (see methods; REFS), the populations of larger consumers are more resilient to the effects of starvation, whereas those of smaller consumers are more prone.

% PARAMETERS INFLUENCING SLOPE ONLY
The consumer's maximal rate of reproduction $\lambda^{\rm max}_C$ influences only the mass-density slope except for the case $\chi \rightarrow -1$, where growth becomes zero.
Above this trivial limit, we observe the consumer growth rate to have a negative effect on the mass-density slope, such that as the growth rate increases, the mass-density relationship becomes steeper (Figs. \ref{fig:corr1},\ref{fig:corr2}).
As the intercept does not change, this means that the steady states of larger bodied consumers decline with increasing $\lambda_C^{\rm max}$, while those of smaller-bodied consumers remain unaltered, though the effect is slight.
% Because larger organisms more efficiently convert resource to 
% , though this effect is nearly imperceptible and not of sufficient magnitude to change the mass-density relationship.
Of more interest is the effect of the yield coefficient $Y_C$ and starvation rate $\sigma^{\rm max}$ (Figs. \ref{fig:corr1},\ref{fig:corr2}).
The yield coefficient represents the conversion of resources to consumer biomass, where an increase in $\chi$ correlates to large increases in consumer steady state without altering the mass-density slope. 
Here we observe that increased efficiency in converting resource to consumer biomass will have an effect similar to increasing resource productivity, as the effective abundance of the resource is greater when relatively fewer resources fuel a given unit of consumer biomass.
Because $Y_C \propto E_d$, where $E_d$ is the energy density of the resource (see methods), resource quality is therefore expected to translate directly to higher consumer steady state densities.

\section{Mortality from predation}
\setcounter{figure}{0}
% - explain how the new predation mortality works

\noindent {\bf Per-capita mortality rate from predation} The per-capita mortality rate from predation of the herbivore consumer with mass $M_C$ and population density $C$ by a mammalian predator with body mass $M_P$ and population density $P$ is given by
\begin{equation}
\beta(C,P)=w\frac{\lambda_P(C) P}{C Y_P}, 
\end{equation}
where $\lambda_P(C)$ is the growth rate of the predator, $Y_P$ is the predator yield coefficient, describing the grams of predator produced per gram of prey consumed, and $w$ is the degree of predation intensity ($w=1$ denotes high predation intensity, whereas $w<1$ denotes lower predation intensity).
Assuming a linear functional response for predation mortality, $\lambda_P(C)$ is maximized to $\lambda_P^{\rm max}$ when the consumer reaches its theoretical maximum population density, which we calculate by converting the resource carrying capacity directly to grams of consumer produced, or $C^{\rm max} = Y_C k$.
The growth rate of the predator is then given by
\begin{equation}
    \lambda_P(C) = \lambda_P^{\rm max}\frac{C}{C^{\rm max}} = \lambda_P^{\rm max}\frac{C}{Y_C k}.
    \label{eq:predfunction}
\end{equation}
% where $\lambda_P^{\rm max}$ is the maximum predator growth rate.
Together, we observe the per-capita mortality rate to be (as expected) independent of the consumer density $C$, and is simplified to
\begin{equation}
\beta(P)=w\frac{\lambda_P^{\rm max} P}{Y_P Y_C k}, 
\end{equation}
where we assume that the predator population remains at empirically measured steady state densities for mammalian carnivores, where $P\equiv P^*= P_0M_P^{-0.88}$ \citep{carbone2002common}. 
This assumption is required because the effects of predation are implicit rather than explicit, and effectively assumes that predator populations operating far below this relationship are not viable.
While there is bound to be a range of viable densities for a predator of a given body size, that mass-density relationships exist at all indicates that population densities are highly constrained over evolutionary time and therefore represent a predator energetic demand as a function of body size.
% By assuming predator populations are constrained to this mass-density relationship, they effectively operate on a different timescale than their prey.
Accordingly, if the predator mass-density relationship $P^*$ represents an expected energetic requirement for a functioning predator population, our assumption of predation as a constant, rather than dynamic, influence on herbivore mortality reveals the dynamic consequence of such energetic relationships.
We suggest that it is these energetic mismatches that may constrain longer-timescale macroevolutionary forces, even if the shorter-timescale ecological dynamics may be more idiosyncratic and complex than our minimal model captures.\\

\noindent {\bf Herbivore and predator yields} As described in the main text, consumer yield is calculated
\begin{equation}
    Y_C = \frac{M_C E_d}{\int_0^{t_{\lambda_C}} B_0 m(t)^\eta {\rm dt}},
\end{equation}
where $E_d$ is the energy density of the plant resource $R$ (Joules/g) and the denominator is the lifetime energy use required by the herbivore consumer to reach maturity (Joules).
The parameters $t_{\lambda_C}$ and $B_0$ are the timescale associated with reaching reproductive maturity and the metabolic coefficient for herbivorous mammals, respectively, and $\eta=-3/4$ is the metabolic exponent (see Table 1, main text).
The predator yield is calculated similarly, where
\begin{equation}
    Y_P = \frac{M_C E_C}{\int_0^{t_{\lambda_P}} {B_{0}}_P m(t)^\eta {\rm dt}},
\end{equation}
where $E_C$ is the energy density of the herbivore being consumed, and the denominator is the lifetime energy use required by the predator to reach maturity.
The parameters $t_{\lambda_P}$ and ${B_{0}}_P$ are the timescale associated with reaching reproductive maturity and the metabolic coefficient for predatory mammals, respectively, and $\eta=-3/4$ is the metabolic exponent.
We note that the metabolic coefficient for predators is different than that for mammals \citep{MunozGarcia2005}.

% \noindent {\bf Herbivore energy density} 
The energy density of herbivore consumers changes with body mass $M_C$.
For example, small mammals have very low percent body fat, whereas very large mammals have high percent body fat.
We assume that predators consume all non-skeletal mass of prey. 
Because the amount of consumable tissues with different energy densities within an herbivore varies allometrically, so too should the energy density $E_C$.
We consider four primary tissue groups: a consumable set composed of muscle, fat, and \emph{other} tissues, and an non-consumable set composed only of skeletal tissues.
If the scalings associated with fat, muscle, and skeletal tissues are $M_C^{\rm fat}=f_0 M_C^{1.19}$, $M_C^{\rm musc} = g_0 M_C^{1.00}$, and $M_C^{\rm skel} = h_0 M_C^{1.09}$ \citep[with normalization constants $f_0 = 0.02$, $g_0 = 0.38$, and $h_0 = 0.0335$;][]{prange1979scaling}, the scaling of the \emph{other} tissue (gut tissue, organ tissue, etc) is given by $M_C^{\rm other} = M_C - (M_C^{\rm fat} + M_C^{\rm musc} + M_C^{\rm skel})$.
The energy density of fat is $E_{\rm fat} = 37700$ J/g, whereas the energy density of muscle is $E_{\rm musc} = 17900$ J/g \citep{merrill1973part}.
If we assume that gut and organ tissues have roughly the same energy density as muscle, the attainable energy density for an herbivore of size $M_C$ is given by
\begin{equation}
	E_C(M_C) = E_{\rm fat}\frac{M_C^{\rm fat}}{M_C} + E_{\rm musc}\left(\frac{M_C^{\rm musc}}{M_C} + \frac{M_C^{\rm other}}{M_C} \right).
\end{equation} \\

\noindent {\bf Large-bodied Predator-Prey Mass Relationship (PPMR)} The predator growth rate $\lambda_P^{\rm max}$, the time required for the predator to reach reproductive maturity $t_{\lambda_P}$, and the predator's steady state population density $P^*$ are allometric relationships that depend on predator body mass $M_P$.
Accordingly, for an herbivore of a given mass $M_C$, we must anticipate the size of its likely predator $M_P$.
This is very different than the predator-centric perspective of anticipating the average prey size for a given predator \citep{Carbone:1999ju}.
For example, the most preferred prey mass for an African lion is ca. $350$ kg \citep{hayward2005lion}, where the inclusion of megaherbivores to diet is comparatively low.
However from a megaherbivore's perspective, lions may represent the only potential predator.
In other words, because the range of prey body mass increases for predators of larger body mass \citep{Sinclair2003}, it is the upper limit of the range that impacts the populations of larger herbivores.
% To make matters more complicated, 

To obtain an herbivore-centric measure of the expected predator mass given a particular herbivore mass ${\rm E}\{M_P | M_C\}$, we first compiled the known diets of large-bodied predators, including tigers, lions, hyenas, leopards, dhole, wild dogs, and cheetahs \citep{hayward2005lion,Hayward2006hyena,hayward2006leopard,hayward2006lycaon,hayward2006cheetah,Hayward2008}.
Because smaller mammalian predators and prey have very different PPMRs than larger-bodied mammalian predators and prey, we here focus exclusively on the predators of large-bodied herbivore prey $>10^5$ g. 
From the mean proportional reliance of predators on large-bodied prey \citep{hayward2005lion,Hayward2006hyena,hayward2006leopard,hayward2006lycaon,hayward2006cheetah,Hayward2008}, we repeatedly sampled predator dietary distributions to reflect each predator's reliance as a function of prey mass.
We introduced variability in predator and prey masses by assuming that body sizes were normally distributed about the expected value with a standard deviation of $\pm 25\%$, allowing us to obtain a distribution of expected predator diets as a function of prey mass.
From this relationship, we then evaluated the expected predator mass for a given prey mass range to obtain ${\rm E}\{M_P\}$ (Fig. 4b, main text), demonstrating the allometric relationship of ${\rm E}\{M_P\}=9.76\times10^3 M_C^{0.21}$, where we used the output of 100 independent replicates to robustly estimate the best fit.
% 9.76\times10^3 M_C^{0.21}
We emphasize that this relationship only pertains to large-bodied predators and prey $>10^5$ g.
Alterations to and variations from this relationship are explored in the main text.

\begin{figure*}
  \centering
  \includegraphics[width=1\textwidth]{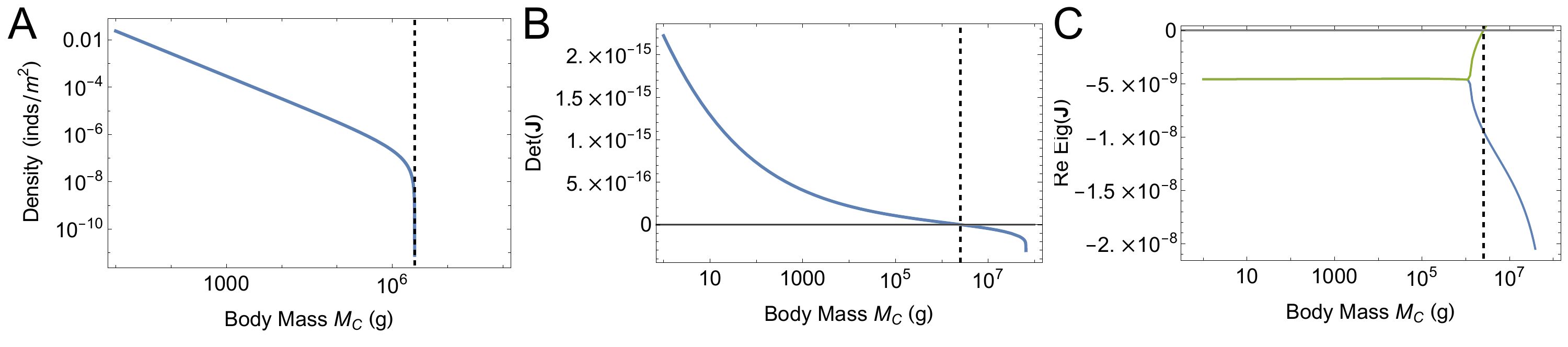}
  \caption{
  \footnotesize{
	  A. Consumer population density as a function of mass $M_C$.
   B. The determinant of the Jacobian matrix for the system presented in eq. 3 (main text) with predation mortality included (eq. XX, main text) as a function of mass $M_C$.
   C. The Real component of the two eigenvalues of the Jacobian assessed in panel B. as a function of mass $M_C$.
   The dashed line in each panel is the measured consumer mass threshold $M_C^\dagger = 2.5*10^6$ g.
   The Gray horizontal line in panels B,C denotes zero on the y-axis.
   }
  }
  \label{fig:stability}
\end{figure*}

Including the empirically-measured PPMR (or a variant of the empirically measured PPMR -- see below) results in the appearance of a transcritical bifurcation at consumer mass $M_C^\dagger$.
We observe that this critical mass threshold results in the extinction of the consumer population characterized by body sizes $M_C \geq M_C^\dagger$ (fig. \ref{fig:stability}A).
At this body mass, the Determinant of the Jacobian matrix characterizing the system presented in eq. 3 (main text) with predation mortality included (eq. 8, main text) is zero (fig. \ref{fig:stability}B), aligning with the real component of a single eigenvalue crossing zero and becoming positive (fig. \ref{fig:stability}C).
While we do not derive a normal form for this bifurcation, these features strongly suggest the observed bifurcation is transcritical in nature \citep{Kuznetsov1995}.

\begin{figure}
  \centering
  \includegraphics[width=1\linewidth]{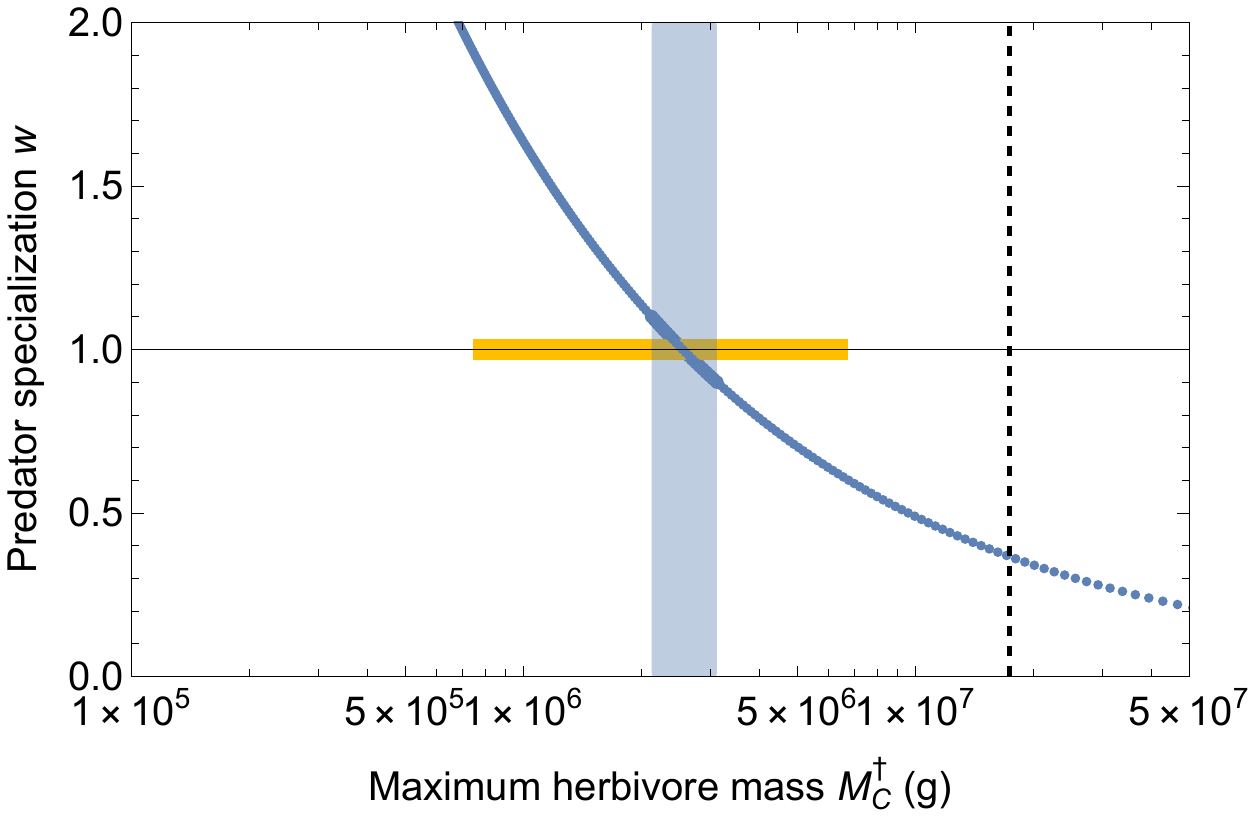}
  \caption{
  \footnotesize{
	  The effect of changing the predation intensity $w$ on the single herbivore consumer population. If $w=1$, predation intensity is maximized, is lower if $0<w<1$, and is above the maximal level required to support a predator population if $w>1$.
	  The blue region denotes herbivore threshold mass range characterizing $w=1 \pm 0.1$.
	  The yellow line denotes the mass range of contemporary elephantids.
	  Vertical dashed line denotes the size of the largest terrestrial mammal (\emph{Deinotherium} at ca. $1.74\times 10^7$, corresponding to $w=0.37$, such that predation intensity is moderate.
   }
  }
  \label{fig:maxpreyvar}
\end{figure}

As explored in the main text, the empirically-measured PPMR for large-bodied mammals results in a threshold body size for herbivore consumers $M_C^\dagger$.
This size marks the point where the predator population, with a body mass derived from the PPMR, cannot sustain its own growth from the predated herbivore population, thereby driving the herbivore population to extinction.
The size at which $M_C^\dagger$ occurs is both dependent on the nature of the PPMR, as well as predation intensity $w$.
As predation intensity $w$ decreases, $M_C^\dagger$ increases (Fig. \ref{fig:maxpreyvar}).

\begin{figure}
  \centering
  \includegraphics[width=1\linewidth]{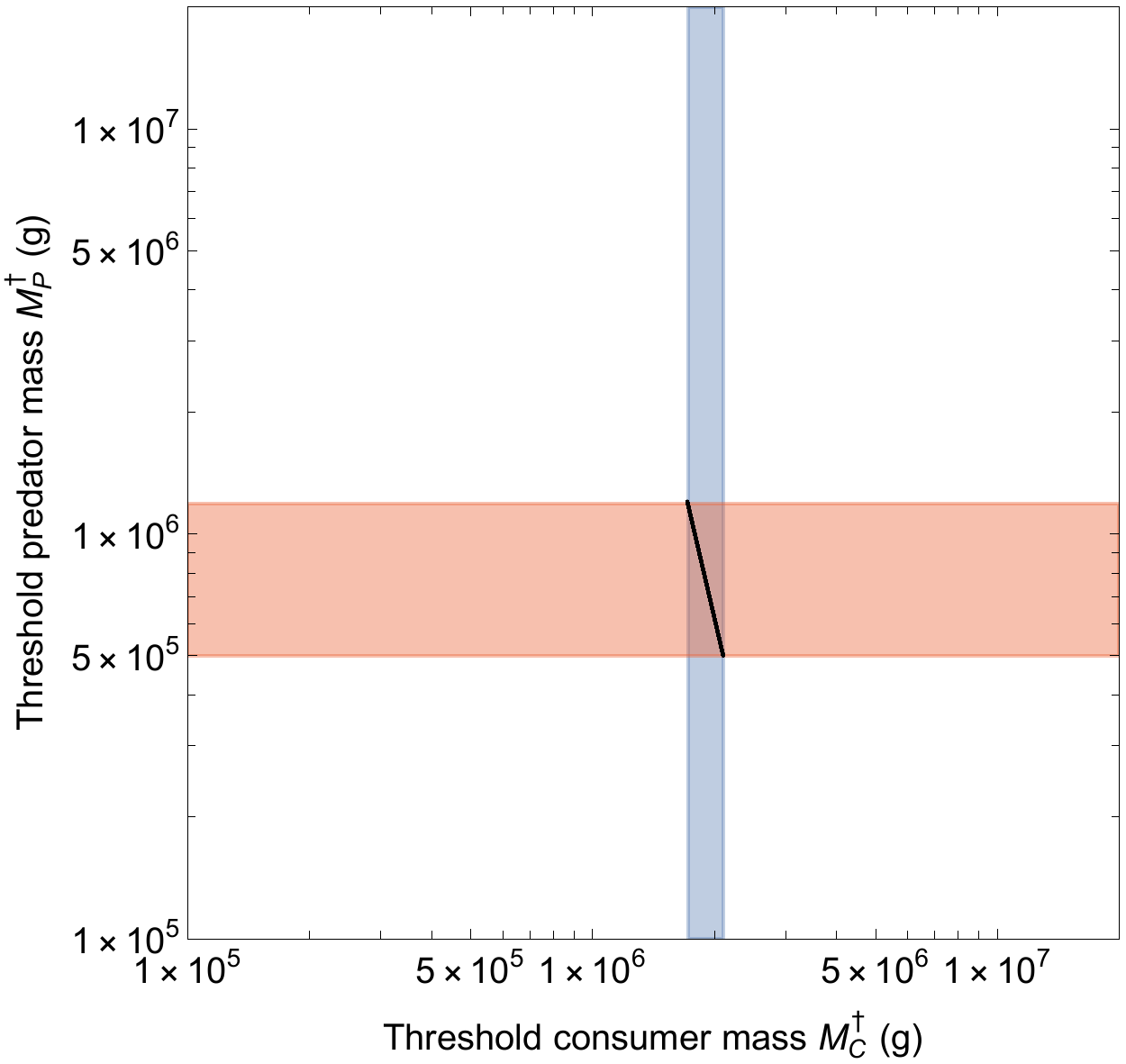}
  \caption{
  \footnotesize{
    Mass ranges corresponding to feasible megatrophic interactions (where herbivore and predator threshold masses are $> 5\times10^5$ g) across variations to the assumed PPMR, demarcated by the white bands in Fig. 4C,D (main text), and assuming high predation intensity ($w=1$). 
% 	  The range of herbivore and carnivore threshold size classes where both species are characterized by mega- body mass ($>6\times10^5$ g) under the condition of predator specialization and across variable PPMRs, where ${\rm E}(M_P|M_C) = p_0(1+\chi_{\rm int}) M_C^{p_1(1+\chi_{\rm slope})}$ and both $\chi_{\rm int}$ and $\chi_{\rm slope} \in (-0.99,2)$.
}
  }
  \label{fig:megamammal}
\end{figure}

\begin{figure*}
  \centering
  \includegraphics[width=1\textwidth]{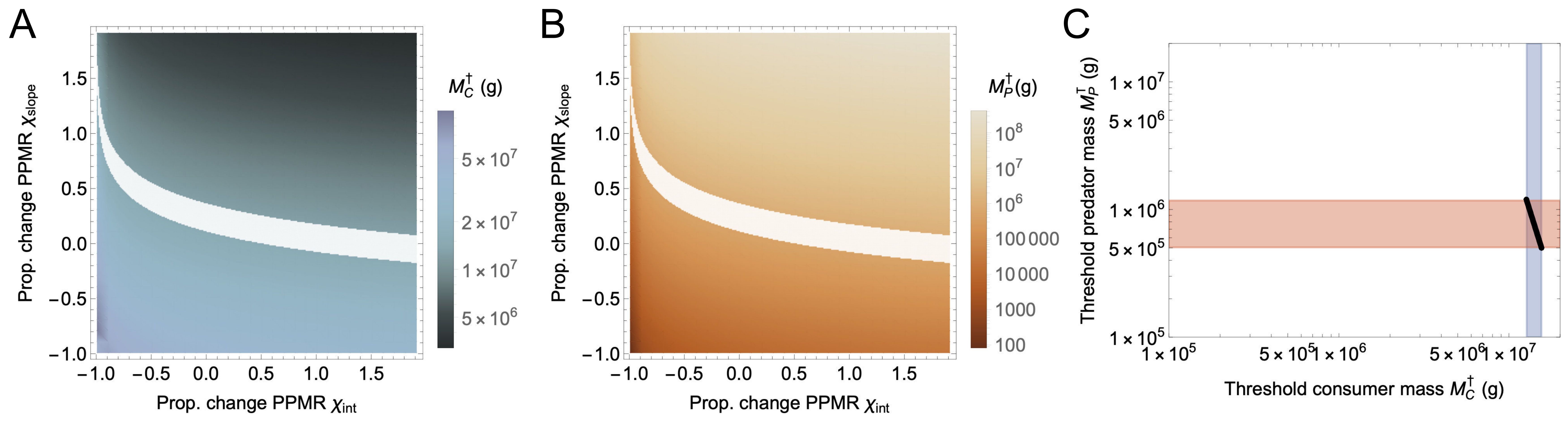}
  \caption{
  \footnotesize{
    The effects of lower predation intensity ($w=0.37$) on A. threshold herbivore mass $M_C^\dagger$ and B. threshold predator mass $M_P^\dagger$ across variable PPMRs, where ${\rm E}\{M_P\} = v_0(1+\chi_{\rm int}) M_C^{v_1(1+\chi_{\rm slope})}$ and both $\chi_{\rm int}$ and $\chi_{\rm slope} \in (-0.99,2)$, and $v_0 = 9.76\times10^3$ and $v_1 = 0.21$ are set as in the main text.
    White bands denote regions of $\chi_{\rm int}$ and $\chi_{\rm slope}$ where megatrophic interactions are feasible (i.e. both predator and herbivore threshold masses are $> 5\times10^5$ g).
    C. Mass ranges corresponding to feasible megatrophic interactions in the white bands in A. and B.
    }
  }
  \label{fig:megamammalgen}
\end{figure*}

By allowing the PPMR to vary as 
\begin{equation}
{\rm E}\{M_P\} = v_0(1+\chi_{\rm int}) M_C^{v_1(1+\chi_{\rm slope})},
\label{eq:ppmr}
\end{equation}
where the proportional changes in the PPMR intercept and slope are given by $\chi_{\rm int}$ and $\chi_{\rm slope} \in (-0.99,2)$,
so does the threshold herbivore body mass $M_C^\dagger$ and, by extension, the related threshold predator body mass $M_P^\dagger$.
From Fig. 4D (main text), we observe that changing the intercept and slope of the PPMR has a large influence on $M_C^\dagger$ and $M_P^\dagger$.
Across this range of potential PPMRs, we highlight those values for the intercept and slope of the PPMR that permit megatrophic interactions, where both megapredators subsist on megaherbivores at the threshold body mass (highlighted region in Fig. 4D, main text).
Fig. \ref{fig:megamammal} shows the relationship between megapredator and megaherbivore body masses highlighted within this region.
Allowing both the PPMR to vary and assuming lower predation intensity ($w=0.37$) enables much larger body sizes for megaherbivores and their associated megapredators (Fig. \ref{fig:megamammalgen}).

\begin{figure}
  \centering
  \includegraphics[width=1\linewidth]{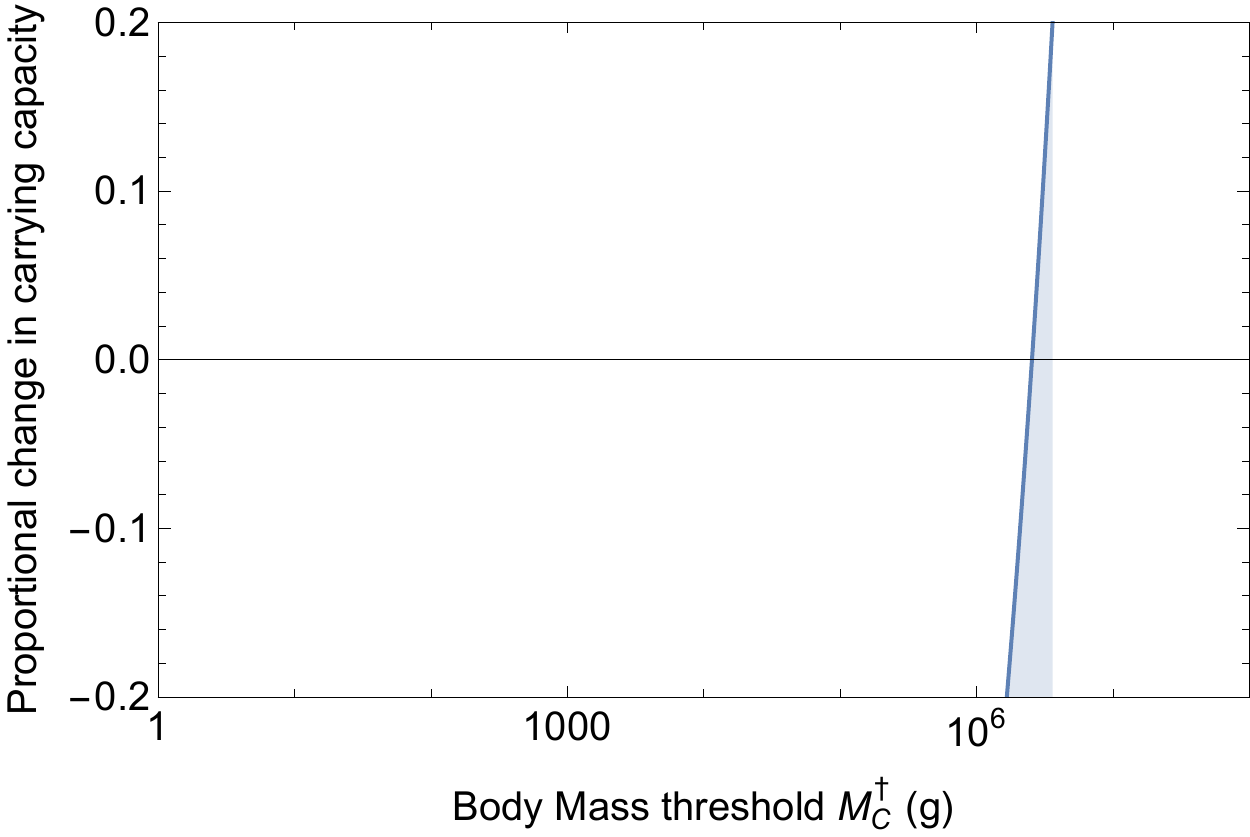}
  \caption{
  \footnotesize{
    The effect of an altered carrying capacity $k$ on the consumer mass threshold $M_C^\dagger$.
    Carrying capacity values that vary from -20\% to +20\% give rise to an $M_C^\dagger$ roughly 65\% and 140\% the original estimated mass of $2.55\times10^6$ g.
    Equivalent variations in $\alpha$ will result in the same changes to $M_C^\dagger$ given its role in eq. 7 (main text).
    The shaded region highlights the mass range captured by the altered carrying capacity.
    }
  }
  \label{fig:carryingcapacity}
\end{figure}

Finally, we note that changes to both the resource growth rate $\alpha$ as well as to the resource carrying capacity $k$ can impact the consumer size at which populations become infeasible $M_C^\dagger$.
Because the consumer steady state is directly proportional to both of these parameters (see eq. 7, main text), a lower growth rate and/or carrying capacity lowers the intercept of the steady state mass-density relationship $C^*(M_C)$.
Analysis of the effect of changes to $k$ (and this will be the same for $\alpha$) reveals that while it has influence on $M_C^\dagger$, it is not incredibly large (Fig. \ref{fig:carryingcapacity}.
However this relationship carries with it an important message: in environments with lower carrying capacities and/or plant growth rates, we would expect a lower mass threshold bounding feasible megaherbivore populations.

\section{Derivation of harvesting mortality}
\setcounter{figure}{0}
% - explain how the new harvesting mortality works
We first determined the harvest rate $h=h^\dagger$ required to drive an herbivore population to extinction, thereby satisfying the condition $C^*(M_C|h) = 0$ as a function of herbivore body mass $M_C$.
This extinction-inducing harvest rate, itself now a function of consumer body mass $h^\dagger (M_C)$, defines the rate at which the population must be harvested to drive the steady state to zero.
To compare this rate against measures of harvest both in nature and predicted from other mathematical or computational treatments of harvest-induced extinction, we  calculated the harvest pressure $\psi^\dagger$, which we defined as the number of herbivore individuals per area harvested at this rate to reduce the population to some proportion $\epsilon$ of its steady state.
This harvest pressure is thus defined by some number of individuals harvested per year over a certain number of years to reduce the population from $C^*$ to its post-harvest density $\epsilon C^*$.

To calculate harvest pressure, we first assume that at the steady state, harvest is occurring on a shorter-than-generational timescale.
For megaherbivores such as elephants, a generation is roughly 25 years \citep{wittemyer2013comparative}, and for harvest pressures that must be applied beyond this period of time, we would expect population growth to counter the negative effects of harvest.
Assuming harvest-only change, we simplify the dynamics to
\begin{equation}
    \frac{\rm d}{\rm dt}C = -h^\dagger (M_C) C,
\end{equation}
where the time to reduce $C^*$ to $\epsilon C^*$ is
\begin{align}
    C(t) &= C_0{\rm e}^{-h^\dagger (M_C) t}, \nonumber \\
    \epsilon C^* &= C^*{\rm e}^{-h^\dagger (M_C) t} \nonumber \\
    t_\epsilon &= -\frac{\log(\epsilon)}{h^\dagger(M_C)}.
\end{align}
We note that for elephant-sized herbivores and larger, $t_\epsilon \leq 23$ years.
While the time required to harvest the population to $\epsilon C^*$ is only just approaching generational timescales, it should be treated as a minimum $t_\epsilon$ given the effects of population growth will prolong the imposed harvest effort. 
Harvest pressure is then calculated as
\begin{equation}
    \psi^\dagger = \frac{C^*(1-\epsilon)}{M_C t_\epsilon}c_0 = -h^\dagger (M_C) \frac{C^*(1-\epsilon)}{M_C \log(\epsilon)}c_0
\end{equation}
where the constant $c_0$ denotes the conversion from inds/m${}^2$/second to inds/$A_{\rm CA}$/year, where $A_{\rm CA}=4.24\times10^{11}~{\rm m}^2$ is the arbitrarily-chosen area of California.
This conversion is particularly important for evaluating other harvest measures from the historical record and estimates from independent models and simulations for extinct species.
As described in the main text, the extinction-inducing harvest pressure is calculated to be $4.3\times10^3$ inds/yr/$A_{\rm CA}$ for an elephant-sized organism of $M_C = 2.5\times10^6$ g (see Fig 5, main text).\\

\noindent {\bf Harvest pressure on Pleistocene mammoths} We compare our measure of harvest pressure to that calculated for mammoths (\emph{Mammuthus primigenius}) by \citet{fordham2022process}.
Because \citet{fordham2022process} employ a much more specific and detailed assessment of the effects of harvest specifically for mammoths over a spatially explicit landscape, we must make a few simplifications in order to derive a comparable estimate.
First, the harvest interaction between mammoth populations and humans is modeled as a Type II functional response, where, again isolating population-level effects to that of harvest we obtain
\begin{equation}
    \frac{\rm d}{\rm dt}C = -\frac{sNFC}{G + \frac{C}{C_{\rm max} M_C}},
\end{equation}
where $N$ is the normalized human population density maximized at unity, the constant $s=7.884\times10^{-8}$ generations/second (where a generation is 25 years), 
$F$ represents the effectiveness of human hunting, ranging from $(0.01,0.34)$, $C_{\rm max} = 1.875\times10^{-6}$ g/${\rm m^2}$ is the maximum mammoth population density (converted from the average degree-by-degree grid cells in Siberia), $G=0.4$ is the half-saturation constant, and $M_C = 2.5\times10^6$ grams.

Solving for the time required to reduce the population to $\epsilon C^*$, we obtain
\begin{equation}
    t_\epsilon^{\rm mammoth} = \frac{C^* - C_{\rm max}G M_C \log\left[C^*\exp(\frac{C^*\epsilon}{C_{\rm max} G M_C})\epsilon\right]}{sC_{\rm max}F M_C N}.
\end{equation}
% where $\mathcal{W}$ is the Lambert, or product logarithm function.
We then calculate the harvest pressure as
\begin{equation}
    \psi^{\rm mammoth}=c_0\frac{sC_{\rm max} F N (C^*-C^*\epsilon)}{C^*-C_{\rm max}G M_C \log\left[C^*\exp(\frac{C^*\epsilon}{C_{\rm max} G M_C})\epsilon\right]},
\end{equation}
where the constant $c_0$ again denotes the conversion from inds/m${}^2$/second to inds/$A_{\rm CA}$/year, where $A_{\rm CA}$ is the arbitrarily-chosen area of California.
Given a range in $F\in(0.01,0.35)$ and $N\in(0.01,1)$, we obtain a distribution of values for mammoth harvest pressure with a median value of $1.24\times10^4$ inds/yr/$A_{\rm CA}$ over the course of 9.8 years.
The bounds of the estimated range from $5\times10^4$ inds/yr/$A_{\rm CA}$ over the course of 2 years to $5\times10^2$ inds/yr/$A_{\rm CA}$ over the course of ca. 200 years (the range is plotted as the vertical black line in Fig. 5, main text).\\
% Again we emphasize that these calculations of harvest pressure are derived from an extinction timescale that should be viewed as a minimum estimate given that we do not account for demographic rebound.\\

\noindent {\bf Harvest pressure on Pleistocene \emph{Diprotodon}} The harvest rate needed to collapse \emph{Diprotodon} populations was calculated by Bradshaw et al. \citep{bradshaw2021relative}, where a harvest pressure of between 400-500 inds/year/area of Australia was sufficient to collapse the population.
Translating this to the area of California, we obtain between 678 to 848 inds/yr/$A_{\rm CA}$, with a mean of $763.2$ inds/yr/$A_{\rm CA}$.\\

\noindent {\bf Harvest pressure on historical elephants \emph{Loxodonta africana}} Elephant harvest rates are estimated from historical documentation of the ivory trade detailed in \citet{milner1993exploitation}.
While the trade volume oscillates with changes in technology, access to habitats within Africa, and the feedbacks of trade on elephant population size, we compare our results against estimates taken at two points in time: early in the ivory trade (1810), and late in the ivory trade (1987). 
From \citet{milner1993exploitation} we assume that each elephant killed contributes 1.88 tusks, and that tusk mass begins at 15 kg per tusk early in trade to 5 kg per tusk in later years.
While the area from which elephants were harvested is largely unknown, we assume the area harvested is that assessed to be suitable elephant habitat in sub-Saharan Africa, estimated at $3.22\times10^{12}~{\rm m}^2$ \citep{thouless2016african}.
From rates of ca. $1\times10^5$ kg/yr of ivory harvested in 1810 to ca. $9.7\times10^5$ kg/yr of ivory harvested in 1987, normalized to habitat area and converted to the area of California, we obtain estimates of ca. $467$ inds/yr/$A_{\rm CA}$ in 1810 to ca. $1.33\times10^4$ inds/yr/$A_{\rm CA}$ in 1987 (see Fig. 5, main text).

% \clearpage
% \subsection{Supplementary Figures}

% \begin{figure*}
%   \centering
% \includegraphics[width=0.9\textwidth]{harvest_scaling_plot.png}
% \caption{
%   }
%   \label{fig:harvestscaling}
% \end{figure*}

\clearpage

\pagebreak

%%%%%%%%%% Insert bibliography here %%%%%%%%%%%%%%
% \bibliographystyle{amnatnat}
% \bibliography{aa_starving3}

% \putbib

\end{bibunit}

\end{document}